\documentclass[fleqn]{statsoc}

\usepackage{natbib}
\usepackage[a4paper]{geometry}

\usepackage[utf8]{inputenc}
\usepackage[T1]{fontenc}

\usepackage{graphicx, caption}
\usepackage{subcaption}

\usepackage[tbtags]{amsmath}
\usepackage{amsfonts,amssymb}
\DeclareMathSizes{13}{18}{14}{10} 

\usepackage{comment}
\usepackage{lineno}
\usepackage{hyperref}
\usepackage[explicit]{titlesec}
\usepackage{longtable}
\usepackage{booktabs}
\usepackage{multirow}
\usepackage{makecell}

\usepackage{hyperref}
\usepackage[nameinlink]{cleveref}
\Crefname{figure}{Figure}{Figures}
\Crefname{equation}{Eq.}{Eqs.}

\newcommand{\be}{\begin{equation}}
\newcommand{\ee}{\end{equation}} 
\newcommand{\bea}{\begin{eqnarray}}
\newcommand{\eea}{\end{eqnarray}}

\newcommand{\e}{\mathrm{e}}

\newcommand{\Gam}{\textrm{Gam}}
\newcommand{\multi}{\textrm{Mult}}
\newcommand{\cat}{\textrm{Cat}}

\newcommand{\pois}{\textrm{Pois}}

\renewcommand{\bf}[1]{\textbf{#1}} 

\newcommand{\f}[2]{\frac{#1}{#2}}

\newcommand{\ccup}[1]{\left\{#1\right\}}

\newcommand{\bup}[1]{\left(#1\right)}
\newcommand{\rup}[1]{\left[#1\right]}

\newcommand{\Exp}{\mathbb{E}}

\renewcommand{\ref}[1]{[\ref{#1}]}

\usepackage[algoruled]{algorithm2e}
\usepackage{algorithmic}
\usepackage{sidecap}
\usepackage{boldline}
\usepackage{setspace}

\graphicspath{{./figures/}}

\newcommand{\vimure}{\mbox{{\small VIMuRe}}}

\title[Latent Network Models to Account for Noisy, Multiply-Reported Social Network Data]{Latent Network Models to Account for Noisy, Multiply-Reported Social Network Data}

\author{Caterina De Bacco}
\address{Max Planck Institute for Intelligent Systems, Cyber Valley, Tuebingen 72076, Germany}
\email{caterina.debacco@tuebingen.mpg.de}

\author{Martina Contisciani}
\address{Max Planck Institute for Intelligent Systems, Cyber Valley, Tuebingen 72076, Germany}

\author{Jonathan Cardoso-Silva}
\address{Department of Methodology, London School of Economics and Political Science, Houghton Street, London, WC2A 2AE, United Kingdom}

\author{Hadiseh Safdari}
\address{Max Planck Institute for Intelligent Systems, Cyber Valley, Tuebingen 72076, Germany}

\author{Diego Baptista}
\address{Max Planck Institute for Intelligent Systems, Cyber Valley, Tuebingen 72076, Germany}

\author{Gabriela Lima Borges}
\address{Department of Human Behaviour, Ecology \& Culture, Max Planck Institute for Evolutionary Anthropology, Leipzig 04103, Germany}

\author{Tracy Sweet}
\address{Department of Human Development and Quantitative Methodology, University of Maryland, College Park, MD 20742}

\author{Jean-Gabriel Young}
\address{Department of Mathematics and Statistics and Vermont Complex Systems Center, University of Vermont, Burlington, VT 05405}

\author{Jeremy Koster}
\address{Department of Anthropology, University of Cincinnati, Cincinnati, OH 45221, USA}
\address{National Science Foundation, Alexandria, VA 22314, USA}

\author{Cody T. Ross}
\address{Department of Human Behaviour, Ecology \& Culture, Max Planck Institute for Evolutionary Anthropology, Leipzig 04103, Germany}

\author{Richard McElreath}
\address{Department of Human Behaviour, Ecology \& Culture, Max Planck Institute for Evolutionary Anthropology, Leipzig 04103, Germany}

\author{Daniel Redhead} 
\address{Department of Human Behaviour, Ecology \& Culture, Max Planck Institute for Evolutionary Anthropology, Leipzig 04103, Germany}
\email{daniel\_redhead@eva.mpg.de}

\author[De Bacco {\it et al.}]{Eleanor A. Power}
\address{Department of Methodology, London School of Economics and Political Science, Houghton Street, London, WC2A 2AE, United Kingdom}
\email{e.a.power@lse.ac.uk}

\begin{document}

\begin{abstract}
  Social network data are often constructed by incorporating reports from multiple individuals. However, it is not obvious how to reconcile discordant responses from individuals. There may be particular risks with multiply-reported data if people's responses reflect normative expectations---such as an expectation of balanced, reciprocal relationships. Here, we propose a probabilistic model that incorporates ties reported by multiple individuals to estimate the unobserved network structure. In addition to estimating a parameter for each reporter that is related to their tendency of over- or under-reporting relationships, the model explicitly incorporates a term for ``mutuality,'' the tendency to report ties in both directions involving the same alter. Our model's algorithmic implementation is based on variational inference, which makes it efficient and scalable to large systems. We apply our model to data from 75 Indian villages collected with a name-generator design, and a Nicaraguan community collected with a roster-based design. We observe strong evidence of ``mutuality'' in both datasets, and find that this value varies by relationship type. Consequently, our model estimates networks with reciprocity values that are substantially different than those resulting from standard deterministic aggregation approaches, demonstrating the need to consider such issues when gathering, constructing, and analysing survey-based network data. 
\end{abstract}

\section{Introduction}

Social network analysis has emerged as a fruitful framework for social scientists to represent and understand social relationships and their consequences~\citep{borgatti_network_2009}. 
For example, patterns of interaction among people, as well as individuals' perceptions of their relationships, have been found to be important for their material wealth~\citep{jackson_inequalitys_2021}, social position and welfare~\citep{lin2002social, redhead2021hierarchy}, and health and well-being~\citep{perkins_social_2015, holt_loneliness_2015}. 

While new data sources now allow for the study of digitally-mediated interactions~\citep[such as social media, mobile phone records, and other trace data;][]{eagle_inferring_2009, park_strength_2018, lazer2021meaningful}, social scientists' interest in day-to-day interactions and interpersonal relations are not always amenable to direct observation.
Researchers therefore continue to rely on surveys where respondents identify the people with whom they have interactions or social relationships~\citep{burt_network_1984}. 
A variety of approaches exist for eliciting self-reported network ties from respondents. 
Most common is the ``name generator'' method, where respondents are asked to list the names of those with whom they have different types of relationships or interactions. 
Other approaches require a full roster, where respondents are asked about their relationship(s) with a set of possible partners~\citep{warner_new_1979, marsden_recent_2005, ross_dietryin_2021}. 

Importantly, survey-based elicitations can be used not only for accounts of concrete interactions or exchanges, but can also facilitate a representation of respondents' subjective perceptions of their connections \citep{krackhardt_cognitive_1987, freeman_filling_1992}. 
Questions may be framed around more qualitative sentiments towards others---such as friendships---and so do not merely document concrete interactions or observed events of exchange. 
For many substantive research questions, an individual's imperfect perception of their social relationships may be as (if not more) important as observable events of interaction or exchange. 
This has been highlighted by empirical research suggesting that individuals place considerable weight on their subjective relationships when making important decisions about who to cooperate with or support~\citep{power_social_2017, redhead2021coalitions, von_rueden_dynamics_2019}, and by work demonstrating that such relationships have strong associations with many important social and health-related outcomes~\citep{kristiansen2004social, smith2008social}. 

The applicability of self-reported network data, however, has been subject to enduring debate within the social networks literature. 
Particularly when prompts query concrete exchanges or interactions, the quality of such data rests on the reliability of the self-reports that respondents provide, and numerous empirical studies have highlighted a plethora of potential biases in responses~\citep{killworth_informant_1976, bernard_problem_1984}.
There is evidence that respondents' recall of their ties can be low, even over short periods of time~\citep{brewer2000forgetting}.
For example, women within two West African communities were only able to accurately recall between $53\%-59\%$ of their interactions across a 24 hour period prior to surveying~\citep{adams_measuring_2006}.
Alongside this, individual differences in the ability to recall ties may be predicted by relationship type, the number of partners, and the duration of a given relationship~\citep{bell_partner_2007}.
Both theoretical studies and empirically observed patterns of nomination also suggest that individuals expressing particular attributes~\citep[e.g., high social status or power;][]{simpson2011power} are more readily named, regardless of whether a relationship actually exists~\citep{ball2013friendship, marin_are_2004, redhead2021reliable, shakya_exploratory_2017, marineau_individuals_2018}.
The order in which questions appear within a survey, and the mode of elicitation, may further influence responses~\citep{pustejovsky_question-order_2009, eagle_methodological_2015}.
That is, respondents have been shown to become fatigued, and report fewer relationships, when asked several name generator questions~\citep{yousefi2019relationship}. 
And, responses can also vary between interviewers, based in part on their attributes and their dynamic with the interviewee~\citep{marsden_interviewer_2003, lungeanu_using_2021}. 

Noting all of these potential biases, one common practice is to obtain multiple reports on any single tie within a network.
For relationships that are understood to be undirected, this is inherently captured with a single name generator question (i.e., both members of a friendship have the opportunity to report it). 
Previous research has found mixed results as to the concordance between respondents about the existence of their social relationships, with agreement in nominations ranging between $40-90\%$~\citep{adams_tell_2007, marsden_network_1990}. 
For relationships that are understood to be directed, multiple queries are necessary. 
One common approach is to ``double sample'' a relationship, by asking respondents both who they go to for some type of assistance, and also who comes to them~\citep{nolin_food-sharing_2008}. 
When combined with complete sampling, double sampling provides two perspectives on all relationships within the network, as both the giver and receiver have an opportunity to name their partner for one of the two prompts. 
A recent survey of double-sampled network data has suggested that concordance between reporters is low, with an overall average of $10\%$ agreement~\citep{readymeasuring}.
Respondents need not be limited to reporting on the relationships in which they are directly involved, but may also be asked about the relationships between other individuals within the network. 
This type of data has been collected through ``cognitive social structures'' roster designs---where respondents report on the relationships between all individuals within the network~\citep{krackhardt_cognitive_1987, newcomb1961acquaintance}---though respondent fatigue means that this elicitation technique is somewhat uncommon. 
When it has been used, it has also shown relatively low levels of concordance between responses, and has highlighted the individual differences that may guide respondents' perceptions of their relationships and the relationships of others~\citep[see][for a review]{brands2013cognitive}. 

These low levels of concordance suggest that while having multiple reports on any relationship certainly provides new information, it does not necessarily resolve the issue of bias in reporting. 
Indeed, new issues may be introduced, if there are, for example, different reporting propensities for different queries. 
One key issue for double-sampled data in particular may be people's expectation of, or desire for, mutually supportive, balanced relationships~\citep{heider_psychology_1958}.  
The multiple prompts entailed in double sampling may lead to an inflation of observed reciprocity, driven primarily by people's propensity of naming of the same individuals across both prompts~\citep{readymeasuring}, which we here term ``mutuality.'' 
Overall, the low levels of concordance found in multiply-reported data raises the question of how to integrate such imperfect reports elicited from respondents with appropriate statistical models. 

To examine the individual biases that shape self-reports of ties, and estimate the effect of mutuality on core properties of a network, we advance a statistical estimation technique that is able to combine multiply reported network data while accounting for the variable ``reliability'' of respondents.
We validate our model using simulated data, to determine whether we are able to recover the true underlying network, mutuality, and individual reporters' reliabilities. 
Finally, we evaluate our model with two empirical datasets, both with double-sampled questions, one based on a ``name generator'' approach and the other based on the roster method. 
We conclude by discussing our findings and outlining possible extensions to the model.

\subsection{Related Work}

In the social sciences, simple deterministic rules are often used to aggregate potentially  multiple reports on what should nominally be the same relationship \citep{krackhardt_cognitive_1987, lee_mutual_2018}. 
With double sampling, for example, it is sometimes assumed that if one party forgets to report a relationship when asked, (e.g., when they are asked who they give advice to), the other party may report that tie (e.g., when they are asked who they receive advice from); with such an expectation the union of the two name generators is typically used~\citep[e.g.,][]{nolin_food-sharing_2010, ready_why_2018}.
Alternatively, it could be assumed that only relationships that are mutually recognised are salient; with such an expectation the intersection of the two name generators would be the implied choice~\citep[e.g.,][]{krackhardt_friendship_1990}. 
These aggregation rules rest on simple but strong expectations, and presume consistency in how reporters respond to these questions. 
This, paired with the fact that standard statistical tools used in the social sciences (e.g., exponential random graph models \citep{robins2007introduction}, or social relations models \citep{snijders1999social}) assume that reported ties are a ``true'' representation of the network, can potentially lead to serious misrepresentations of the social relations of interest. 

In recent years, several statistical methods have been proposed to resolve discordant reports, in social network analysis \citep{holland_1983_stochastic, killworth_informant_1976, kenny1984social} but also in other fields like systems engineering \citep{amini2004issues}, the biological sciences \citep{sprinzak2003reliable, d2004estimating, hobson2020guide}, and physics~\citep{newman_network_2018}.
Recently, for example, social scientists have attempted to tackle the problem of concordance by computing a ``credibility score'' for every individual within a network, and determining whether a given tie exists base on each reporter's assigned credibility \citep{an_analysis_2015}. 

Among this vast literature, most relevant to the work presented here are Bayesian approaches to network estimation \citep{butts_network_2003}.
They rely on an explicit generative models for the reports, in which the reports are viewed as giving indirect and imperfect information about an unobserved network.
Virtually all of these generative models are formulated as finite mixtures~\citep{titterington1985statistical}, where responses about a particular tie have different distributions depending on the presence or absence of that tie.
Formalized thus, these models can often be estimated with efficient algorithms such as expectation-maximization algorithms \citep{newman2018estimating, newman_network_2018}, exact Gibbs samplers~\citep{butts_network_2003, young2021clustering}, or Hamiltonian Monte Carlo samplers applied to closed marginal distributions~\citep{young_bayesian_2021, young2021reconstruction,redhead2021reliable}.

Generative approaches are fairly flexible and can readily be adapted to diverse experimental and observational situations.
For example, by allowing the distributions of the data to vary from tie to tie, mixture models of networks have been shown to accommodate surveys where respondents have various degrees of trustworthiness \citep{butts_network_2003}, some of the data is missing \citep{peixoto_reconstructing_2018, peixoto2019network} and more~\citep[see][]{redhead2021reliable}.
Further, important social phenomena driving tie formation, such as triadic closure, reciprocity and group structure, can be integrated through specification of priors on the network  \citep[such as with stochastic block models;][]{peixoto_reconstructing_2018, peixoto2021disentangling} or latent space models \citep{butts_network_2003}), or by suitably incorporating them into a likelihood distribution \citep{safdari2021reciprocity,safdari2021reciprocityDyn,contisciani2021JointCrep}.

Given the modelling and computing advantages provided by generative mixture models, we introduce a model complementing the previous work outlined above.
We go beyond existing approaches by introducing a model that simultaneously handles multiply-reported ties and weighted reports, while allowing individuals to vary in reliability.
For technical reasons, the resulting model is no longer a \emph{finite} mixture (unlike previous work), which in turns means most of the standard statistical methods do not apply. 
As we will show, this issue can be resolved by introducing an efficient  variational inference procedures.

\section{The model}

Consider the problem of collecting a network of ties between individuals.
These ties could, for instance, represent relationships commonly studied in the social sciences---such as loaning money, giving advice, or sharing food. 
This can be done by querying a set of $M$ reporters about the existing ties. The real network is not observed; responses of the reporters are the only observed data at our disposal.
We assume that the unobserved network is correlated with these responses. 
Mathematically, we define this as an $ N \times N$-dimensional adjacency matrix, $Y^{}$, where entries $Y^{}_{ij}\in\{0,1\}$ indicate whether a tie  $i \rightarrow j$ exists. 
For each tie type, the observed data is an $N\times N \times M$-dimensional tensor, $X^{}$, with entries $X_{ijm}^{}$ containing reports by respondent $m$ about the tie $i \rightarrow j$. 

We assume that each reporter can, in principle, report on any tie within the network.
The exact rule of how reporters respond may change with the application, but may be flexibly represented by a binary mask $R$ of entries $R^{}_{ijm}$.
We set $R^{}_{ijm}=1$ whenever a reporter $m$ reports a tie from node $i$ to node $j$, and set the entry to 0 otherwise. 
In scenarios where a network has been double-sampled---e.g., where the same reporter responds about \textit{giving} and \textit{receiving} social support---every tie type is sampled twice (for each reporter), one for each direction of the interaction. 

For instance, $m$ can nominate who she gives advice to (giving) and who she goes to for advice (receiving). 
In this case $m \in \ccup{i,j}$ for a tie between $i$ and $j$, and we distinguish the reported data of each tie by $X_{ijm}^{}$ and $X_{jim}^{}$, depending on it being given or received. Generally, we assume the reported data $X$ to be asymmetric in $(i,j)$.
While we gave an example for ties of type \textit{advice}, the  model applies for any type of tie.
 
One of the main objectives of our model is to estimate the structure of the unobserved network, $Y$, from the reported data, $X$. 
We adopt a probabilistic approach where we assume that  $X$ depends on $Y$ in a potentially noisy way. 
This means that we infer a probability distribution over possible structures compatible with the reported ties.
We assume \textit{conditional independence} between the entries of $X$, \textit{given} $Y$ and the model's parameters. 
This is a common assumption made in network models \citep[e.g.,][]{newman_network_2018, peixoto_reconstructing_2018, young_bayesian_2021}, and makes the estimation of the model more tractable. 
Typical exceptions where this assumption may not hold are scenarios where an upper limit is set on the maximum number of nominations a reporter can make---e.g., when respondents are asked ``\emph{Who are your five closest friends?}''. 
In these scenarios there is a (weak) negative correlation between nominations, because the likelihood of nominations reduces each time a nomination is made by a respondent, simply because the respondent is strictly limited to an arbitrary, finite set of nominations \citep{hoff2013fixedRank}.
While this is important to note, this problem is beyond the scope of the current manuscript.

A further core objective for our model is to estimate the reliability of reporters. 
Reporters may under-report (i.e., neglect to report a tie, when it does exist) or over-report (i.e., report a tie, when it does not exist) certain ties, and this may depend on tie type. 
To account for these biased reports we assign a ``reliability'' parameter, $\theta^{}_{m}$, to each reporter, $m$, specific to the tie type given in input. 
For ease of interpretability, we think of this parameter as a positive number taking higher values when the reporter exaggerates their reports and lower values when they under-report. 

Finally, we incorporate the intuition that reporters tend to nominate the same people for both directions of a relationships, $X_{ijm}^{}$ and $X_{jim}^{}$. 
We term this ``mutuality,'' to keep the notion distinct from the standard dyadic reciprocity (henceforth termed reciprocity) in the true unobserved network $Y$. 

Bringing all of these modeling consideration together, we posit that the expected value of the data is:
\be \label{eqn:meancond}
\Exp \rup{X_{ijm}^{}|Y_{ij}^{} = k} = \theta_m\lambda_k^{}+\eta X_{jim}^{}\ .
\ee
where $\eta \geq 0$ is the mutuality parameter.
Mutuality intervenes as an additive and positive contribution to the mean number of reported ties $X_{ijm}^{}$, capturing the possible increasing probability of observing a tie, given that we observed the same tie in the opposite direction, as reported by the same reporter.
The parameter $\lambda^{}_{k}$ is a positive real value that needs to be inferred and regulates the contribution of $Y$ in determining  $X$. 

From this we notice how, for a given set of $\lambda^{}_{k} > 0$, reporters with high $\theta^{}_{m}$ tend to exaggerate their reports, while reporters with smaller values tend to nominate fewer individuals. 
Regardless of the reporter's `reliability', the existence of a tie $X_{jim}^{}$ in one direction increases the probability that the tie $X_{ijm}^{}$ also exists in the opposite direction when $\eta >0$. 
This also implies that it may not be possible to identify the reliability of reporters with high mutuality and in networks with high values of $\eta$. 
In these cases, in fact, the presence of a reported tie can be determined with a high likelihood based on the tie reported on the opposite direction, no further information about the reporter (e.g., reliability) needs to be estimated.  

To form a likelihood for the observed data that can accommodate various network and report structures, in particular directed and weighted reports and networks, we write
\be\label{eqn:L}
P(X_{ijm}^{}|X_{jim}^{},Y_{ij}^{} = k, \lambda_k^{}, \theta_m^{}, \eta) = \frac{(\theta_m^{} \lambda_k^{}+\eta X_{jim}^{})^{X_{ijm}^{}}}{X_{ijm}^{}!} \e^{-(\theta_m^{}\lambda_k^{}+\eta X_{jim}^{})}  \ .
\ee

The model can be applied to any tie type encoded in the input data $X$, and it will output the reliability of a reporter for that tie type. 
One can potentially generalize this to a multi-layer framework by, considering a unique $\theta_m$ for each reporter, regardless of tie type. 
This would then introduce a coupling between the reported $X$ for various tie types, potentially increasing the complexity of the model. 
We do not explore this here.
We assume that there are no contributions to the likelihood of $X$ when a reporter $m$ is not given the chance to report on the tie $i \rightarrow j$.
In empirical applications this could be, for example, when a survey design only asks about ties directly involving the reporter.

In addition to specifying the likelihood as in \Cref{eqn:L}, we adopt a Bayesian approach and assume priors for the parameters and the unobserved $Y$. 
To maximise the flexibility of our model, we allow for different types of values for $X$ and $Y$, leading to the formulation:
\be
P(Y_{ij}^{};p_{ij}^{}) =  \prod_{k}\bup{p_{ij,k}^{}}^{Y_{ij,k}^{}}\ ,
\ee 
with $\sum_{k}p_{ij,k}^{}=1$. 
The resulting model can thus accommodate, for example, a binary network $Y$ and weighted reports $X$. 
We then consider Gamma priors for the remaining parameters, as they are defined for positive real numbers, and they are conjugate with the Poisson distribution, which makes calculations convenient.

\section{Inference}

Because of the possibility of mutuality in nominations we do not have a closed-form joint distribution for $(X_{ijm}^{}, X_{jim}^{})$, hence we use a pseudo-likelihood approximation as in \citet{safdari2021reciprocity}:
\bea
P\bup{\ccup{X_{ijm}^{}}_{m}\vert \ccup{X_{jim}^{}}_{m},Y_{ij}^{}, \lambda^{}, \ccup{\theta_m^{}}_{m}, \eta} = \prod_{m} P(X_{ijm}^{}|X_{jim}^{},Y_{ij}^{}, \lambda^{}, \theta_m^{}, \eta) \nonumber \\
=\prod_{k}\rup{P(Y_{ij}^{}=k) \prod_{m} P(X_{ijm}^{}|X_{jim}^{},Y_{ij}^{}=k,\lambda_k^{}, \theta_m^{}, \eta)}^{Y_{ij,k}^{}} \,.
\eea

The full posterior is then:
\begin{small}
\bea \label{eqn:fullpost}
P(Y,\lambda,\theta, \eta|X) &\propto & P(X|Y, \lambda,\theta, \eta) P(Y) P(\lambda)P(\theta)P(\eta) \nonumber \\
&=& \prod_{,i,j} P\bup{\ccup{X_{ijm}^{}}_{m}\vert \ccup{X_{jim}^{}}_{m},Y_{ij}^{}, \lambda^{}, \ccup{\theta_m^{}}_{m}, \eta} P(Y_{ij}^{};p_{ij}^{}) \\
&& \prod_{,k}P(\lambda_{k}^{}; a_{k},b_{k}) \prod_{,m}P(\theta_{m}^{}; \alpha_{m},\beta_{m}) P(\eta; c,d) \nonumber \\
&=:& \mathcal{L}(\lambda,\theta, \eta, Y) \quad,
\eea
\end{small}
\noindent where the proportionality is due to the omission of an intractable normalization that does not depend on the parameters. 
To estimate the model we use variational inference with a mean-field variational family \citep{blei2017variational}, which finds an approximate posterior distribution for the network and parameters.
The algorithmic updates needed to find the best approximation to the posterior distribution follow a coordinate ascent routine, iteratively finding the best the marginal posterior distribution of each parameter while holding the others fixed. 
We call the resulting algorithm \vimure, for Variational Inference for Multiply-Reported data.
Pseudo-code for the resulting algorithm is shown in Algorithm \ref{alg:cvi}; see \Cref{sec:vi} for further details. 

\begin{algorithm}[!h]
  \setstretch{0.1}
  \caption{{\vimure}}
  \SetAlgoLined
  \DontPrintSemicolon
  \BlankLine
  \KwIn{Data $X$, Model $\mathcal{L}$, Variational family $q$.}
  \BlankLine
  Initialize the variational parameters $\gamma, \, \phi,\,\rho,\nu$ to the priors with a small random offset.
    \BlankLine
  \While{change in {ELBO} is above a threshold}{
    \BlankLine
    For each pair of nodes such that $X_{ijm}^{}>0$, update the multinomials: 
    \BlankLine
    \bea
        \hat{z}^{1}_{mk} &\propto& \exp \ccup{\Psi(\gamma_{m}^{shape})-\log \gamma_{m}^{rate}+\Psi(\phi_{k}^{shape})-\log \phi_{k}^{rate} } \nonumber\\
         \hat{z}^{2}_{ijm}&\propto& \exp \ccup{\Psi(\nu^{shape}) -\log\nu^{rate} + \log X_{jim}^{}}=X_{jim}^{} \exp \ccup{\Psi(\nu^{shape}) -\log\nu^{rate} } \nonumber
        \eea
  where the proportionality is such that $\hat{z}^{1}_{mk} + \hat{z}^{2}_{ijm}=1$.
    \BlankLine
      \BlankLine
 For each reporter, update the reliability parameters:
     \BlankLine
\bea
\gamma_{m}^{shape} &=&\alpha_{m} + \sum_{i,j,k}R_{ijm}^{}\, \rho_{ij,k}^{}\, X_{ijm}^{} \,\hat{z}^{1}_{mk} \nonumber\\
\gamma_{m}^{rate}&=&\beta_{m}+  \sum_{i,j,k}R_{ijm}^{}\, \rho_{ij,k}^{}\f{\phi_{k}^{shape}}{\phi_{k}^{rate}}  \nonumber\ .
\eea
    \BlankLine
 For each possible value $k$ of $Y_{ij}^{}$, update the parameters:
     \BlankLine
\bea
\phi_{k}^{shape} &=&a_{k} + \sum_{i,j,m} R_{ijm}^{}\, \rho_{ij,k}^{}\, X_{ijm}^{}  \,\hat{z}^{1}_{mk}\nonumber\\
\phi_{k}^{rate}&=&b_{k}+  \sum_{i,j,m} R_{ijm}^{}\, \rho_{ij,k}^{}\f{\gamma_{m}^{shape}}{\gamma_{m}^{rate}} \nonumber
\eea
    \BlankLine
and:
\bea
 \rho_{ij,k}^{} &\propto& \exp \left\{\log  p_{ij,k}^{}+\sum_{m} R_{ijm}^{}\,\bup{X_{ijm}^{}\,\hat{z}^{1}_{mk} \Exp_{q(\lambda_k^{})} \rup{\log\lambda_k^{}}}  -\f{\phi_{k}^{shape}}{\phi_{k}^{rate}}\sum_{m}R_{ijm}^{}\, \f{\gamma_{m}^{shape}}{\gamma_{m}^{rate}}   
\right\} \nonumber\ .
\eea
    \BlankLine
     Update the mutuality parameters:
     \bea
     \nu^{shape} &=& c + \sum_{i,j,k,}  \rho_{ij,k}^{} \sum_{m}R_{ijm}^{}\,X_{ijm}^{}  \hat{z}^{2}_{ijm}\nonumber\\
      \nu^{rate} &=& d +  \sum_{i,j,,m} R_{ijm}^{}\, X_{jim}^{} \nonumber \ .
           \eea
    
    \BlankLine
  }
      \BlankLine
  \textbf{Output}: Variational parameters ($\gamma, \, \phi,\,\rho,\nu$).\;
  \label{alg:cvi}
\end{algorithm}

\section{Simulation Experiments}

To validate our model, and study its performance in different regimes we simulate synthetic data that reproduce our analysis scenarios---multiply-reported network data that depend on a latent adjacency matrix---using the model itself.
In detail, we first generate the network $Y$ either with a stochastic block model \citep[SBM,][]{holland_1983_stochastic}, a degree-corrected stochastic block model \citep[DC-SBM,][]{karrer2011stochastic} or a probabilistic model with reciprocity \citep[CRep,][]{safdari2021reciprocity}.
We then generate the observed $X$ given a fixed reliability, mutuality and reciprocity parameters and the generated network, collectively denoted by $\Theta = (Y, \theta, \lambda, \eta)$.
We follow the approach described in \citet{safdari2021reciprocity}, and for each reporter $m$ we draw a pair $(X_{ijm}^{},X_{jim}^{})$ consistently with the joint $P(X_{ijm}^{}, X_{jim}^{} \vert \Theta)$ in a two-step sampling routine where we first generate one of the two reported ties and then the second one given the first, see \Cref{sec:syntheticgeneration} for details.

In these simulations we examine our ability to recover (i) the underlying network $Y$ and (ii) the individual reliabilities $\theta_m^{}$. 
We generated synthetic networks reproducing three different scenarios. 
Two of these scenarios are extreme cases where a fraction of reporters ($\theta_{ratio}$) are tagged to be either (a) over-reporters or (b) under-reporters, while all the others are reliable, i.e., they have $\theta_m^{}=1$ and their $X$ entries are deterministically generated. 
In doing this, we show model performance in difficult cases where the proportion of unreliable reporters are increasing.
The third scenario, (c), is more realistic.
In this setting, we have both over- and under-reports, and we drawn $\theta_m^{}$ from a Gamma distribution, providing a broader range of values. 
We vary the difference between $\lambda_1^{}$ and $\lambda_0^{}$, such that the smaller this difference, the noisier the problem gets, and thus the harder the inference tasks.

In all experiments, we fit two versions of the model: a version with mutuality (\vimure$_T$) and a version without (\vimure$_F$). 
To provide a point of comparison, we also compute two baselines estimates of $Y$: 
(i) the \textit{union}, in which a tie exists if at least one reporter reports that tie; 
(ii) the \textit{intersection}, in which all the reporters of a tie have to agree for the tie to exist. 
These two commonly-used baselines represent the most and least inclusive approaches to integrating multiply-reported data, and so provide reasonable comparisons for \vimure. 

\subsection{Results}

We use the F1-score to assess the ability of our model to recover $Y$ (which is binary in our experiments). 
It is defined as the harmonic mean of precision (fractions of inferred ties that actually exists) and recall (fractions of existing ties found by the method). 
This choice is chiefly motivated by the fact that we have unbalanced data, since much fewer ties than possible tend to exists in empirical network as well as in our experiments.

\begin{figure}[!ht]
	\includegraphics[trim={0.5cm 0 0 0},clip,width=0.9\linewidth]{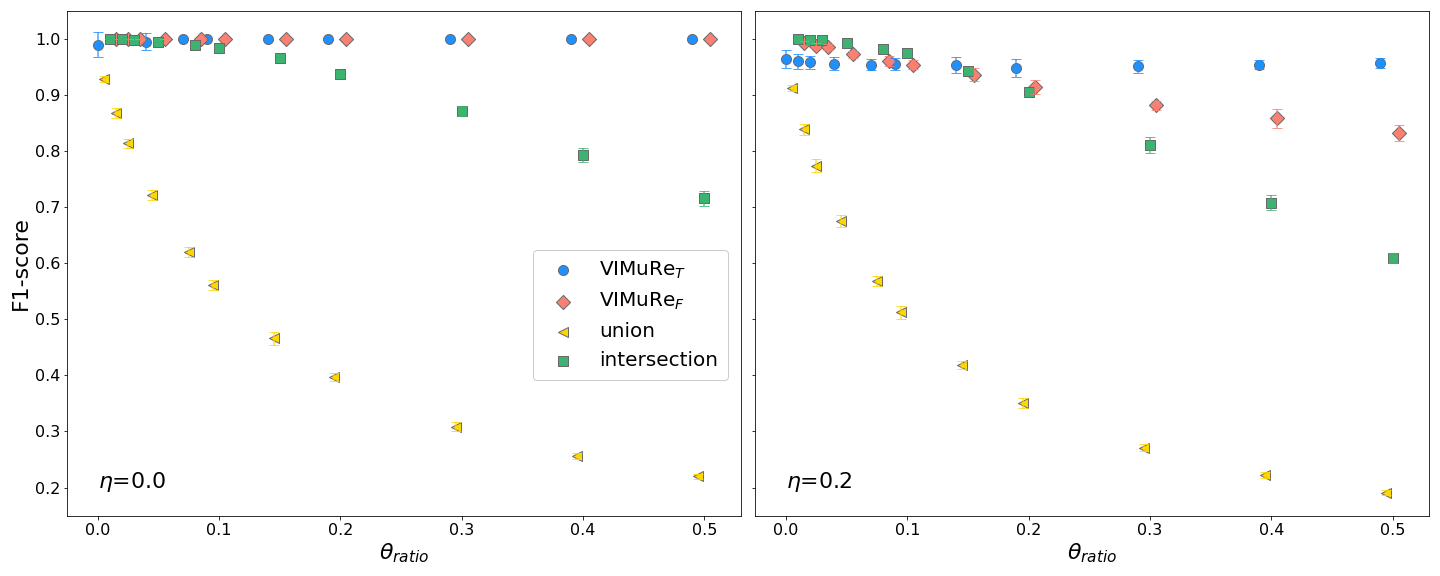}
	\includegraphics[trim={0.5cm 0 0 0},clip,width=0.9\linewidth]{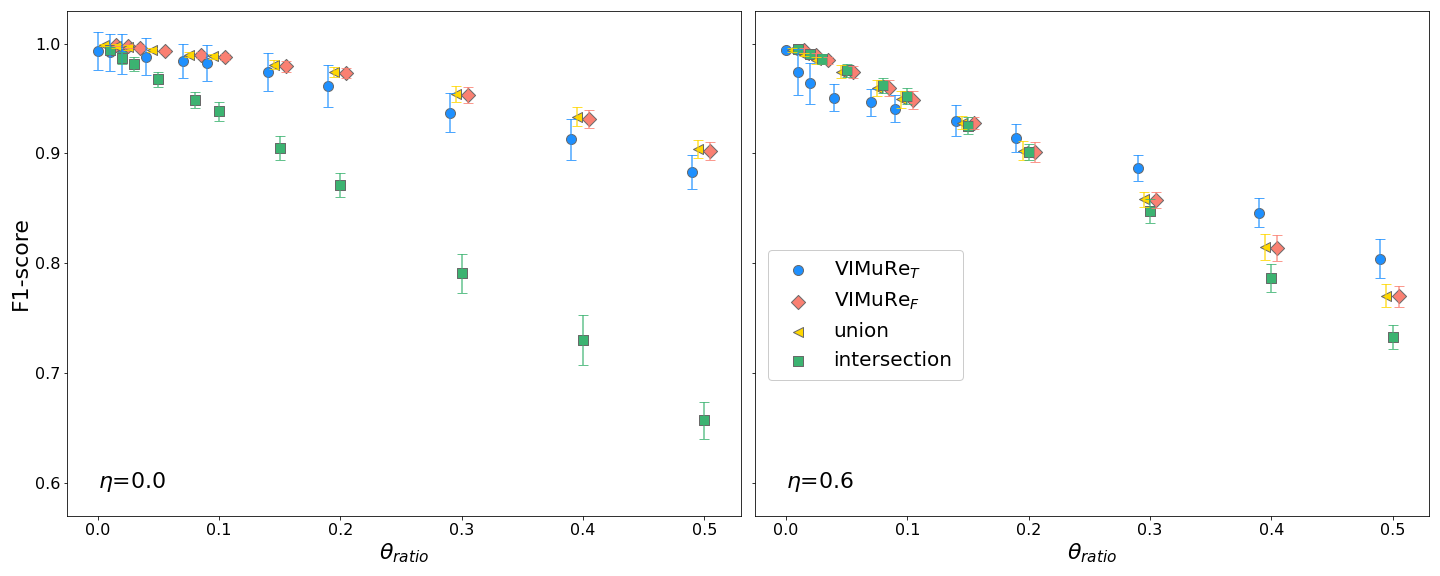}
	\caption{Synthetic networks with over- or under-reporters: estimating underlying network $Y$. Synthetic networks with $N=100$ nodes and $M=100$ reporters generated with the benchmark generative model proposed above by varying the fraction $\theta_{ratio}$ of over-reporters (top) or under-reporters (bottom). The two columns represent networks generated without (left) and with (right) the mutuality effect $\eta$. The results are averages, and standard deviations are calculated over ten independent synthetic networks. The accuracy of the estimate of the underlying network $Y$ is measured with the F1-score. This measure goes from 0 to 1, where 1 indicates perfect matching.}
	\label{fig:synExaUnderY}
\end{figure}  

In the two extreme scenarios, where there are only over- or under-reporters, our model recovers the unobserved network $Y$ better than the union or intersection of $X$. 
The performance of our model is also more robust as the number of unreliable reporters and mutuality increases, see \Cref{fig:synExaUnderY}. 
In particular, our model with mutuality (\vimure$_T$) has a higher performance for high values of $\eta$, which is also a harder regime as the performance of all methods decreases. 
In general, the performance of the baselines decrease as the number of over- or under-reporters grows. 
For example, the union baseline estimates relationships that do not exist in the true network.  when there are several over-reporters.
Conversely, the intersection baseline underestimates the amount of ties when a high fraction of individuals under-report.
Our model overcomes these biases by accounting for reporters' reliability, and this results in higher and more robust performance. 
However, when $\theta_{ratio}$ become too large, \vimure \ also fails since, as \Cref{fig:synExaUnderRel} shows, recovering the reporters' reliability becomes harder.
That said, the model with mutuality performs better at this task, and fails much more slowly than the model without mutuality, especially when $\eta$ is large.

\begin{figure}[htbp]
\centering
\includegraphics[trim={0.5cm 0 0 0},clip, width=0.97\linewidth]{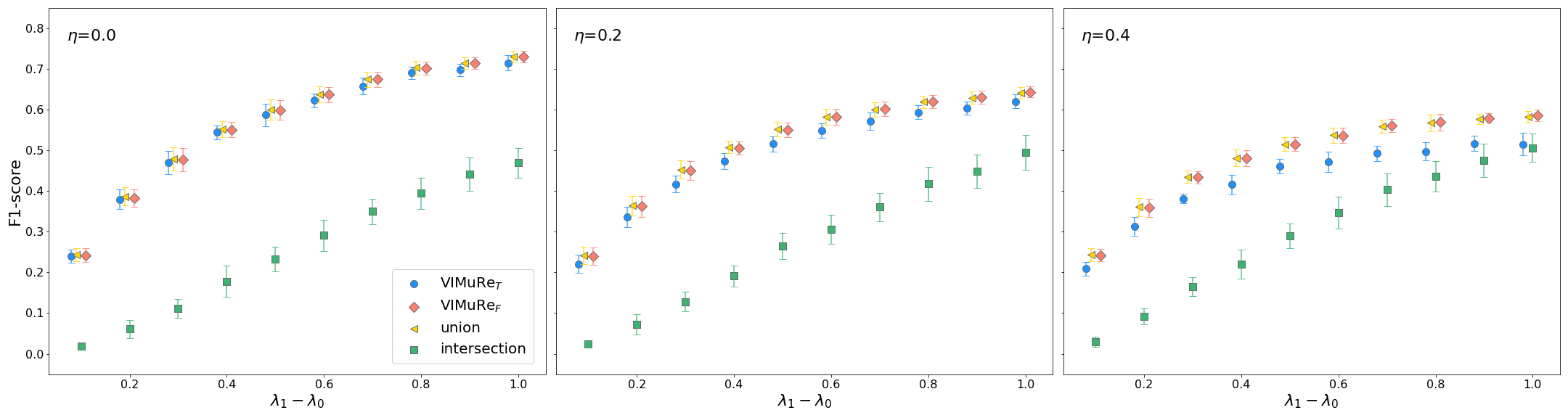}
\caption{Synthetic networks with over- and under-reporters. Synthetic networks with $N=100$ nodes and $M=100$ reporters generated with the benchmark generative model proposed above by varying the difference between $\lambda_{0}$ and $\lambda_{1}$. The three columns represent networks generated with no (left), medium (center) and high (right) mutuality effect $\eta$. The results are averages and standard deviations over twenty independent synthetic networks, and the accuracy over the underlying network $Y$ is measured with F1-score.}
\label{fig:synBothYRel}
\end{figure}

Performance is more nuanced when we consider the more realistic experiment, with a broad range of reporters' reliabilities.
F1 scores are lower than in the previous experiments in general, and recovering the ground truth is particularly harder when the difference between the mean number of reports of a tie being present and not, $\lambda_{1}-\lambda_{0}$, is lower, see \Cref{fig:synBothYRel}.
Intuitively, as the difference $\lambda_{1}-\lambda_{0}$ decreases, both the zero and non-zero inputs of $Y$ tend to make the same contribution in determining $X$; thus, it becomes more difficult to distinguish between reports about ties and absent ties.
These experiments also further confirm what we observed in the previous experiments, that the harder regime is with the highest mutuality. 
A higher $\eta$ means that a reporter will tend to nominate the same set of people for both \textit{giving} and \textit{receiving} questions, which results in having a less structured $X$.
In these experiments, both versions of our model and the union baseline perform similarly while the intersection baseline performs much more poorly. 
However, the performance gap decreases as mutuality increases and more ties are reported.

\begin{figure}[!hbtp]
	\centering
	\includegraphics[trim={0 0 0 0.9cm},clip,width=0.8\textwidth]{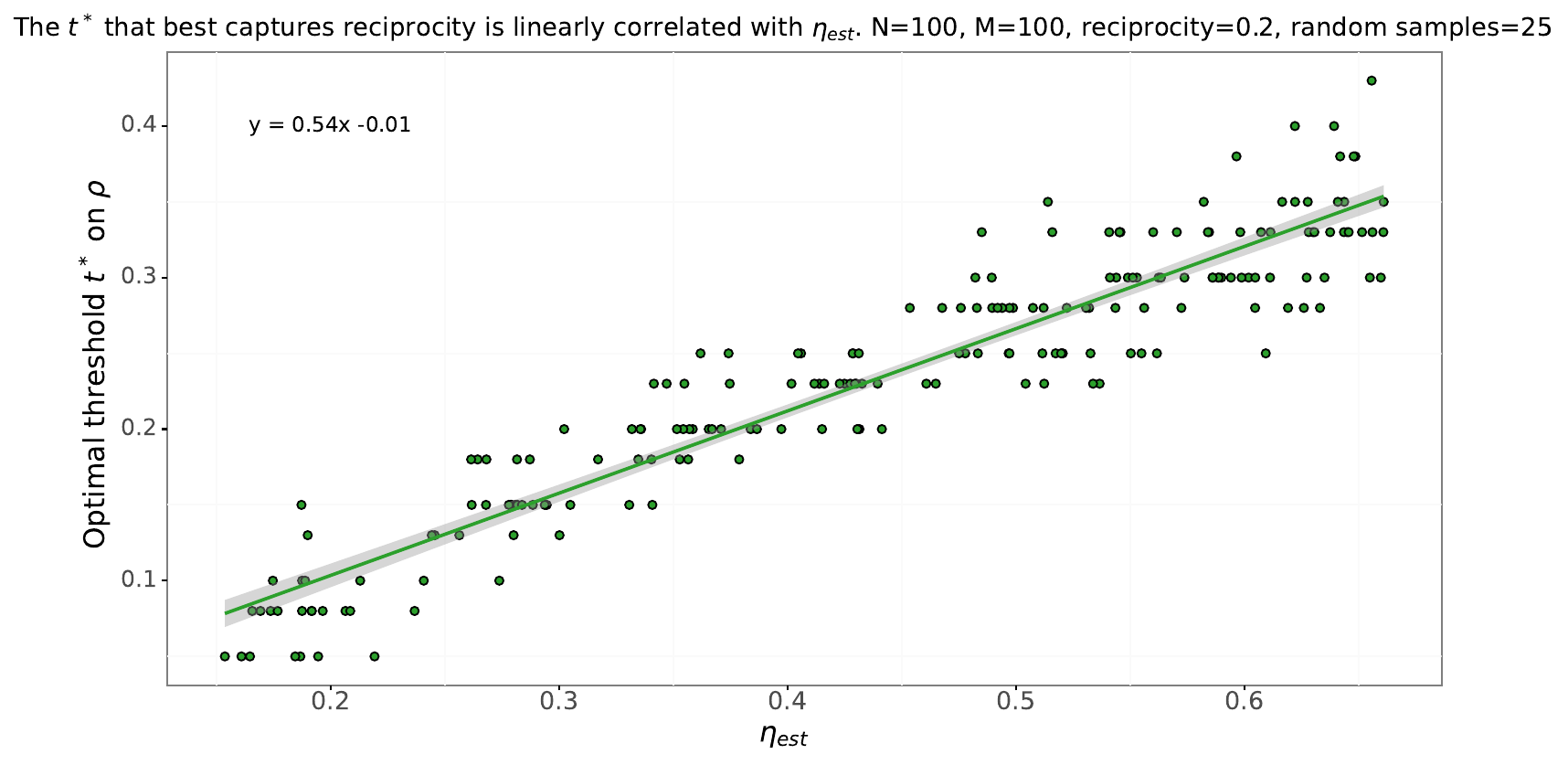}
	\caption{Synthetic networks with $N=100$ nodes and $M=100$ reporters generated with the benchmark generative model proposed above with $\lambda_{1} - \lambda_{0} = 1.0$ and planted reciprocity values around $\approx 0.2$ on the ground truth networks $Y$. The plot shows that the threshold that best captures reciprocity is linearly correlated with $\eta_{est}$.}
	\label{fig:synBestThresholdRho}
\end{figure}

Once we have an estimate $\hat{Y}$ of the unobserved network $Y$, a practitioner would be able to investigate structural properties in the latent network.
To give an example, we will assess the ability of our model to capture reciprocity on the estimated $\hat{Y}$, a foundational feature of many social relations \citep{fehr_reciprocity_1998, molm_structure_2010}.
To this end, we convert the posterior probability distribution $\rho^{}_{ij,k}$ to a binary unweighted adjacency matrix.
Since we considered binary data in our experiments and thus $k \in \ccup{0,1}$, we can obtain this by applying a threshold to the sub-tensor $\rho^{}_{ij,k=1}$, as it represents the probability distribution of finding a $\hat{Y}^{}_{ij}=1$ entry.
After each run of \vimure$_T$, we apply a range of thresholds $t_{\rho} \in [0.050, 0.075, \ldots, 0.725, 0.750]$ such that we assign $\hat{Y}^{}_{ij}=1$ when $\rho^{}_{ij,k=1}\geq t_p$,  and keep track of the best $t^*_{\rho}$, for which reciprocity in the inferred network most closely matched the reciprocity of the ground truth.
In \Cref{fig:synBestThresholdRho} we show that the relationship between this optimal threshold and the mutuality $\eta_{est}$ as inferred by \vimure$_T$ can be approximated by the linear equation: 

\bea
    t^*_{\rho}=0.54~\eta_{est} - 0.01.
\eea

\begin{figure}[!htbp]
\centering
\includegraphics[width=0.77\textwidth]{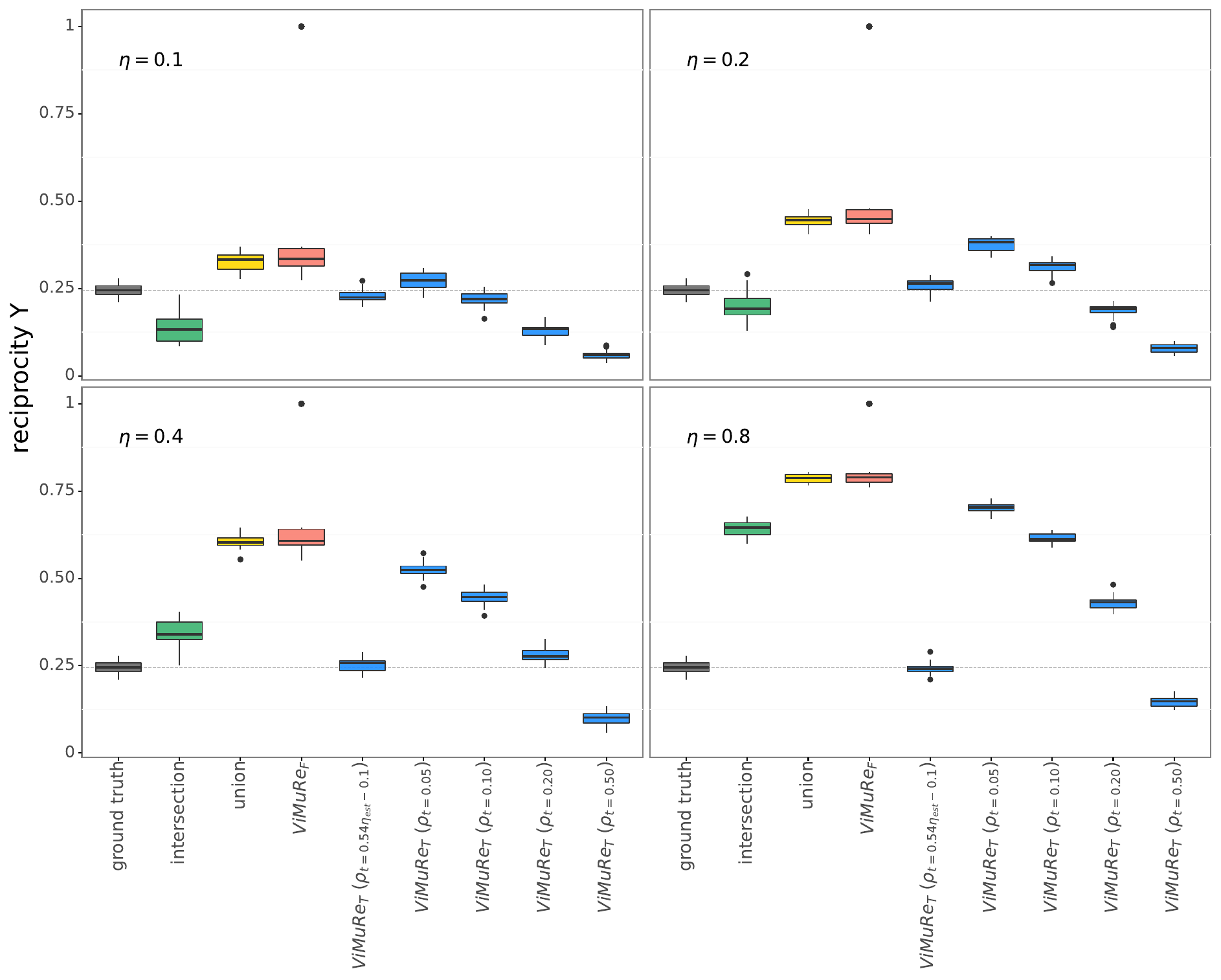}
\caption{Synthetic networks with $N=100$ nodes and $M=100$ reporters generated with the benchmark generative model proposed above with $\lambda_{1}- \lambda_{0} = 1.0$ and planted reciprocity values around $\approx 0.2$ on the ground truth network. The four sub-plots represent networks generated with low (top left), medium (top right) to increasingly high (bottom left and right) mutuality effect $\eta$. The box plots are distributions of the reciprocity on  $Y$, over twenty samples of synthetic networks.}
\label{fig:synBothYRec}
\end{figure}

In fact, in \Cref{fig:synBothYRec} we show that \vimure$_T$ outperforms all other models at this task when the threshold on $\rho$ was set according to the heuristic proposed above.
Reciprocity estimated by the model was a closer match to the reciprocity of the true unobserved network even in simulations with high values of mutuality, a scenario where other methods tend to overestimate reciprocity.
Density of the inferred network is also closer to the ground truth when compared to baseline methods in most scenarios, as it can be seen in \Cref{fig:synBothYDensity}.
These results mean that despite the small gap on the value of F1-score, \vimure$_T$ may be able to provide a better estimate of structural properties on $Y$.

\begin{figure}[htbp]
	\centering
	\begin{subfigure}[b]{0.41\linewidth}
		\includegraphics[width=\linewidth]{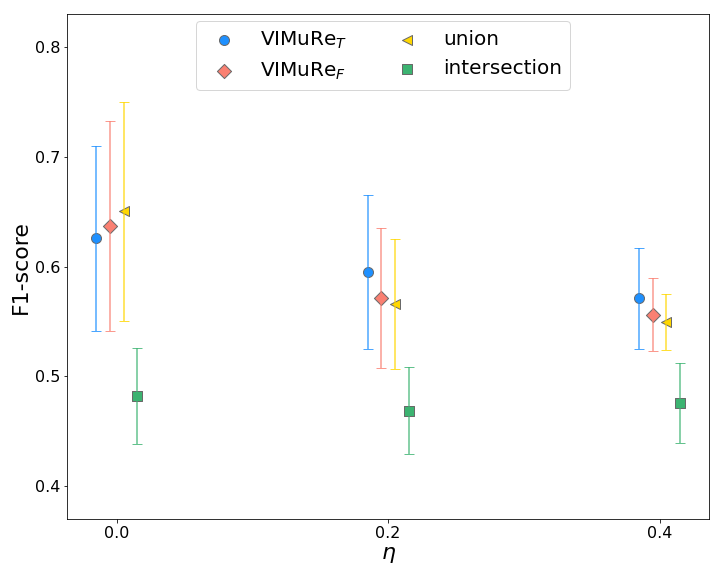}	
	\end{subfigure}
	\begin{subfigure}[b]{0.57\linewidth}
		\includegraphics[width=\linewidth]{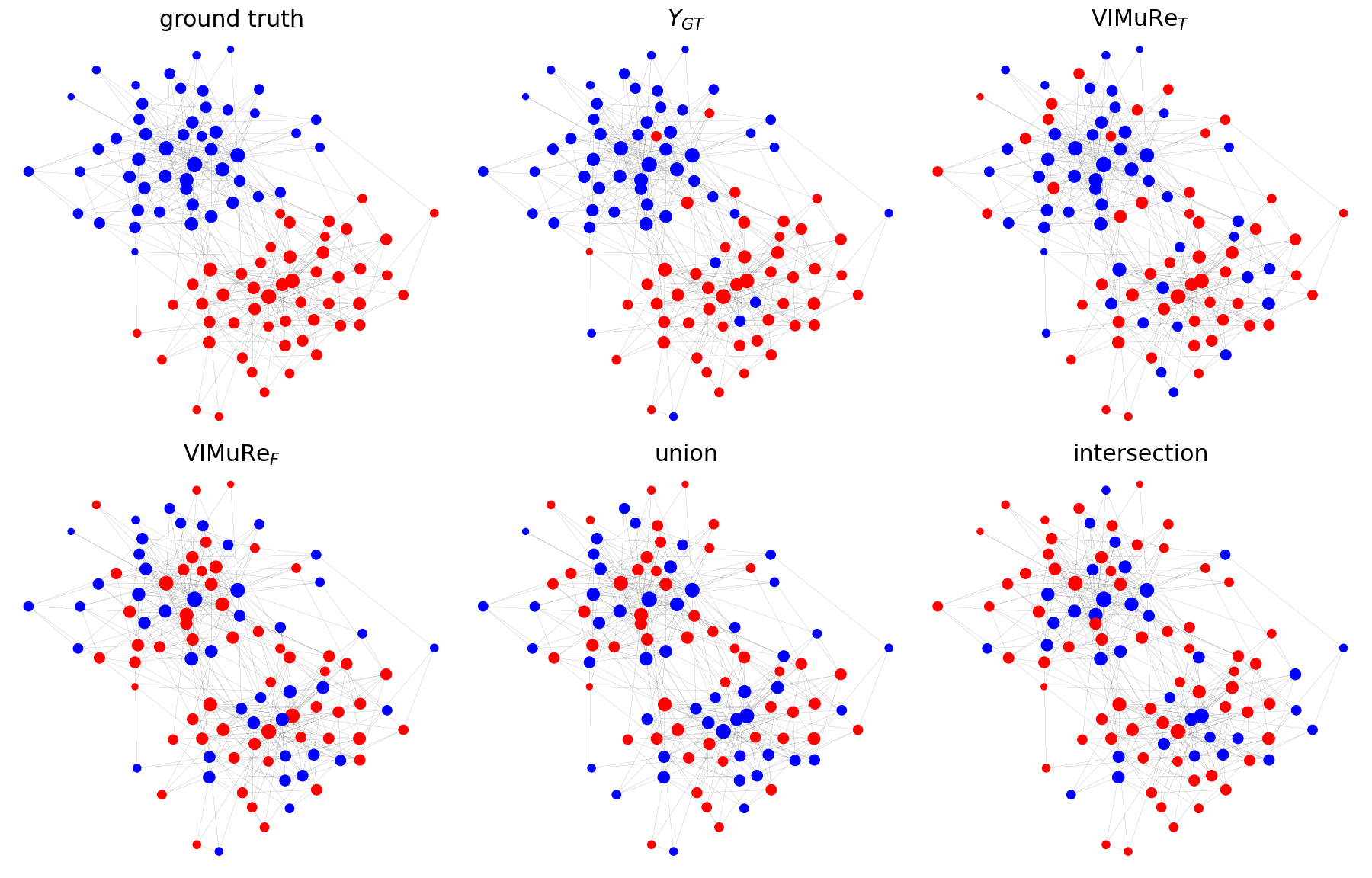}
	\end{subfigure}
	\caption{Synthetic networks  with both over- and under-reporters: recovering community structure. Synthetic networks with $N=100$ nodes and $M=100$ reporters generated with the benchmark generative model proposed above with $\lambda_{1}-\lambda_{0}=1.0$. (left) The results are averages and standard deviations over twenty samples of synthetic networks, generated by varying the mutuality parameter $\eta$. The accuracy in recovering the community structure is measured with the F1-score using the inferred membership vectors. (right) Example of a resulting partition from a synthetic network generated with $\eta=0.4$. The ``ground truth'' is the partition used to generate  $Y$, and $Y_{GT}$ stands for the partition found by using the true $Y$.}
	\label{fig:synBothComm}
\end{figure}

As a final test, we also show that \vimure \ can recover the community structure of a network even when it is measured noisily.
To this end, we use a latent network that has planted communities, and generate reports as before.
We then estimate the network, and finally recover communities using a probabilistic generative model with latent variables \citep{de_bacco_community_2017}, and run it on the estimated $\hat{Y}$ using each method. 
\Cref{fig:synBothComm} shows the result of this experiment and further confirms this in relation to recovering the ground truth communities used to generate $Y$, which is also relevant to many potential applications. 
\vimure \text{} is more robust than the other approaches that have been considered in relation to F1-score across the different mutuality values, providing slightly better results than all other models. 
The intersection performs the worst. 
The qualitative example on the right panel of \Cref{fig:synBothComm} highlights how \vimure$_T$ infers a partition closer to the ground truth than those inferred by the others, where the two true communities are more mixed. 

To summarize, our simulation experiments suggest that working with a generative model with latent variables results in more robust estimates of the underlying and true network, $Y$, in comparison to deterministic baselines (such as union and intersection). 
Furthermore, our model provides an estimate of reporter reliabilities, which can provide additional insights about the networks in and of themselves. 
In addition, we notice better performance of our model when we include the mutuality parameter $\eta$. In particular, \vimure$_T$ shows better results in estimating reciprocity, specifically as people's propensity to report mutuality in their relationships increases. In the application on real-data we use \vimure \text{} with the mutuality parameter. \\

\section{Analysis of social support networks in Karnataka}

We apply our modelling approach to a dataset of social support networks from 75 villages in the Indian state of Karnataka \citep{banerjee_diffusion_2013}.
As part of a larger project, a series of name generators were asked of most of the adult members of a subset of households in each village (overall, about 46\% of all households were surveyed). 
The name generators included questions about four double-sampled relationships: who they would give or receive advice from/to (\textit{Advice}), who they would borrow or lend a small amount of money from/to (\textit{Money}), who they would go to or receive as visitors (\textit{Visit}), and who they would borrow or lend kerosene and rice from/to (\textit{Household Items} -- \textit{``HH Items'' in the plots}).
In the past, these data have been studied by aggregating multiple household respondents answers and taking the union of the double-sampled questions \citep{jackson_social_2012, banerjee_diffusion_2013}.

\begin{figure}[!hbt]
\centering
\includegraphics[width=0.95\linewidth]{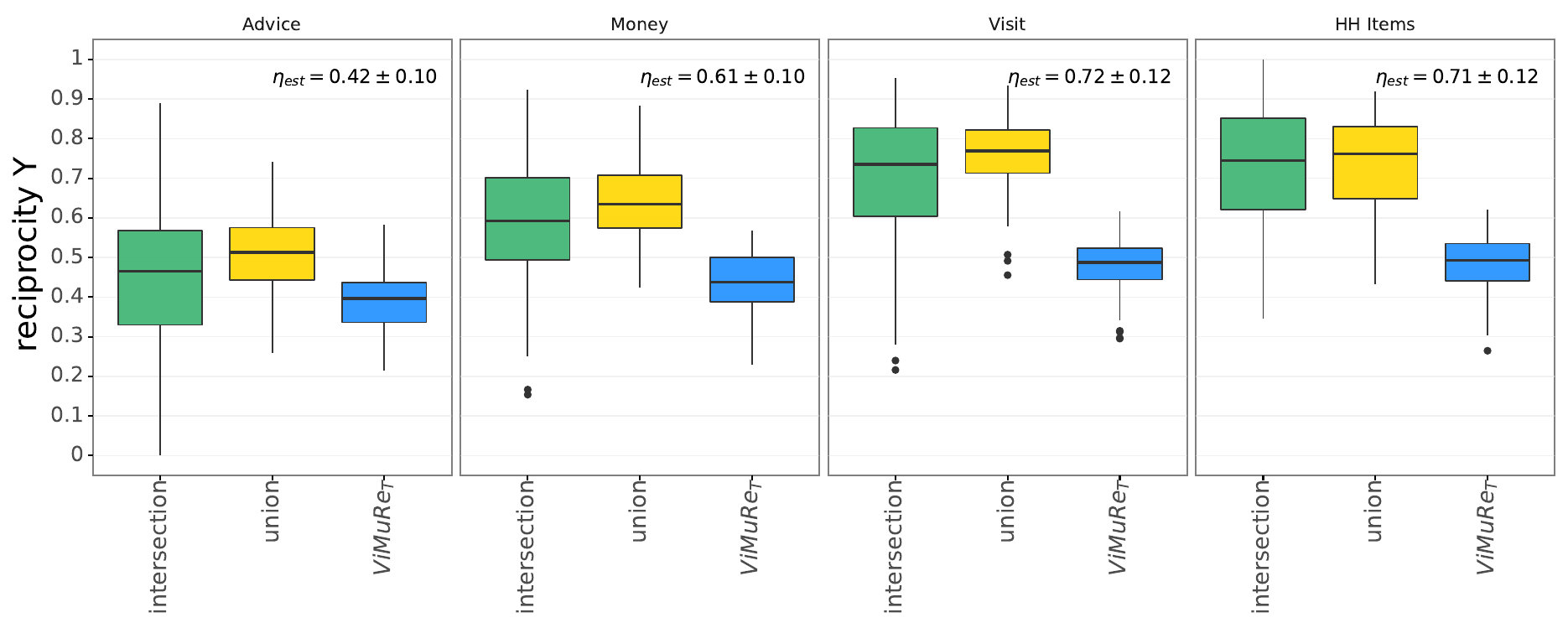}
\caption{Reciprocity on Karnataka networks. The box plots show the distribution over the 75 networks of the reciprocity measured on the inferred $\hat{Y}$. Each column is a different tie type, as written on the figure title.}
\label{fig:recYKarnataka}
\end{figure}  

\begin{figure}
     \centering
     \begin{subfigure}[b]{0.325\textwidth}
         \centering
         \includegraphics[width=\textwidth]{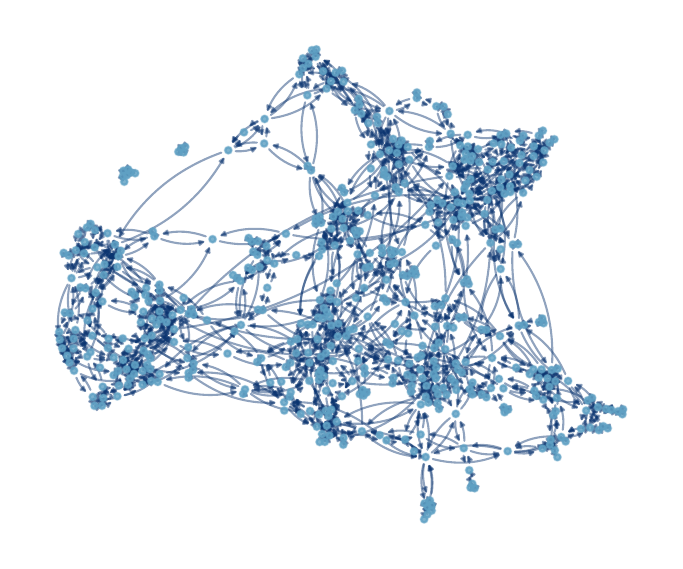}
         \caption{Union ($\text{recip.} = 0.93$)}
         \label{fig:karnataka_networks_union}
     \end{subfigure}
     \hfill
     \begin{subfigure}[b]{0.325\textwidth}
         \centering
         \includegraphics[width=\textwidth]{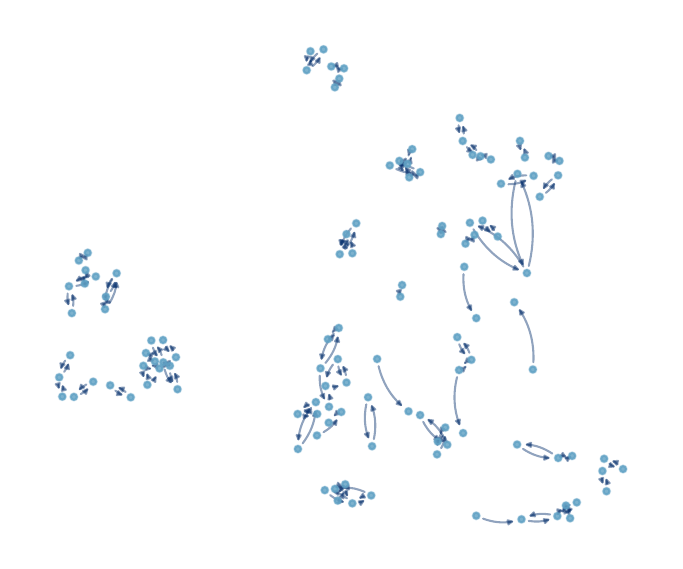}
         \caption{Intersection ($\text{recip.} = 0.88$)}
         \label{fig:karnataka_networks_intersection}
     \end{subfigure}
     \hfill
     \begin{subfigure}[b]{0.325\textwidth}
         \centering
         \includegraphics[width=\textwidth]{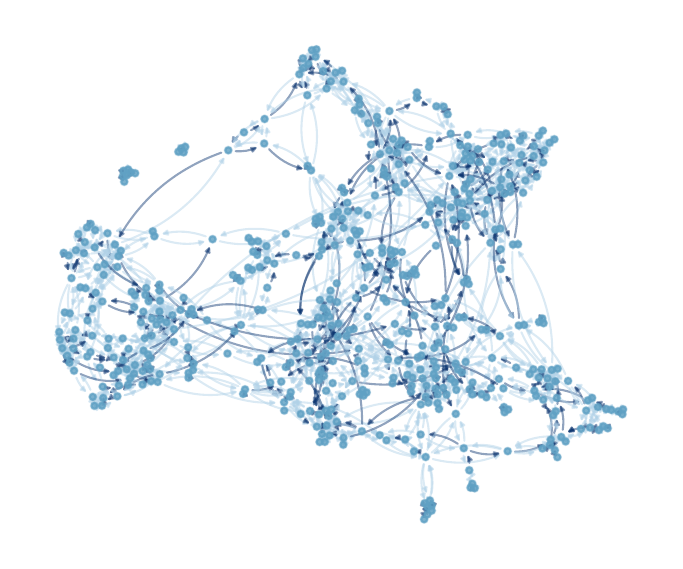}
         \caption{\vimure \text{} ($\text{recip.} = 0.49$)}
         \label{fig:karnataka_networks_vimure}
     \end{subfigure}
        \caption{Example of networks estimated by baseline methods and \vimure \text{} for one Karnataka village (tie type ``Visit''). }
        \label{fig:karnataka_networks}
\end{figure}

Our results suggest that the reciprocity values in these networks are in fact lower than what would be obtained by these simpler approaches, as shown in \Cref{fig:recYKarnataka} and illustrated for one village and tie type in \Cref{fig:karnataka_networks}.  
While we do not have ground truth values in this case, we note that these numbers are similar to those obtained on the synthetic networks in our experiments shown in \Cref{fig:synBothYRec}. 
In particular, they mimic the situation with high mutuality, where the union and intersection significantly overestimate the reciprocity on $Y$, whereas \vimure\text{} identifies the correct range of values.
These results suggest that reciprocity would likely be overestimated in double-sampling network data when the relationships elicited have high mutuality.
Of the four tie types in the data from Karnataka, the estimations made by \vimure\text{} suggest that the ``Advice'' layer has the lowest reciprocity values on average ($0.296 \pm 0.047$), followed by ``Money'' ($0.346 \pm 0.050$), and then the ``Visit'' ($0.367 \pm 0.047$) and ``Household items'' ($0.378 \pm 0.042$) layers exhibiting the highest reciprocity.

Note, too, that our estimates for mutuality ($\eta_{est}$) follow a similar pattern, with the lowest estimates for ``Advice,'' and the highest for ``Visit'' and ``Household Items''. 
These broadly align with expectations: reciprocity---and the \emph{expectation} for reciprocity, as represented by the mutuality term---is higher in those relationships that are understood to be more balanced and mutually supportive (visiting one another's homes and borrowing/lending basic household items like rice and kerosene) and lower in those relationships that are potentially seen as hierarchical and imbalanced (receiving/giving advice and borrowing/lending money). 

We next investigate how reporters' ``reliabilities'' are distributed in these networks. 
Since the mutuality in these graphs is high ($n_{est} \geq 0.4$), it is very common that reporters repeat the same names across the different name generators and therefore 
it is expected that individual ``reliability'' will play a smaller role in determining the latent social network.
Recalling \Cref{eqn:meancond} and the following discussion, this means that the value of $\theta_{m}$ will be small for reporters with a high rate of repeat nominations --- that is, the proportion of alters reported by an ego on the ``give to'' question that gets repeated on the ``gets from'' question. In the Karnataka dataset,  $26\%$ (``Advice'') to $52\%$ (``Household Items'') of reporters have an individual rate of repeated nominations of $100\%$.
These should not be interpreted as under-reporting, as a small $\theta_m$ in this case is just a signal of a predictable reporter's behavior due to high mutuality. 
The vast majority of reporters ($99.49\%$) with a small $\theta_m$ ($\theta_m < 0.1$) in the Karnataka networks have an individual rate of repeat nominations of $100\%$, regardless of the tie type.

For all four tie types we observe that reporters tend to under-report, even after having discounted those individuals who have low $\theta_m$ for the reasons discussed above (see  \Cref{fig:karnatakaReliability}). 
This is consistent with prior literature \citep{butts_network_2003} that suggests that reporters are reasonably accurate when reporting ties, but quite inaccurate when reporting non-ties (as when under-reporting) and may be partially induced by the way the questionnaire was formulated, as there were only four entries available for nominating alters.
This causes under-reporting to be much more likely and substantially limits the number of reported tie configurations that can be observed in this system. 
Indeed, we notice a negative correlation between $\theta_m$ and the in-degree of ties reported by others involving $m$ (see \Cref{fig:indegthetaKarnataka}). 
In particular, reporters nominated by many others (some by more than 20 people) could only nominate up to four among these; such reporters necessarily have low values for $\theta_m$.

We do not observe any strong differences in the distribution of reliability across the four tie types (see \Cref{fig:karnatakaReliability}). 
To explore further, we assess whether reporters are consistent in their reliabilities across the tie types by examining the pairwise distance using the Wassertein distance \citep[a metric for measuring distances between two distributions;][]{Givens1984ACO} between each set of tie types (see \Cref{fig:thetaKarnataka}). 
We see some telling patterns by looking at the consistency of ``Advice'' with the other tie types: while reporters are most consistent between that and ``Money,'' they are least consistent between that and ``Household Items'' and ``Visit.'' 

\section{Analysis of Nicaragua data}

To further highlight the broad applicability and benefits of our modelling approach, we apply \vimure\text{} to data from a horticulturalist community in Nicaragua \citep[see][for more detail on the population and measurement instruments]{koster_family_2018}.
These data were collected through a roster-based design, where all adult residents within the community were presented with a list of all other adult residents and were asked two questions that captured relationships of social support (i.e.,``\emph{Who provides tangible support to you at least once per month?}'' and ``\emph{Who do you provide tangible support at least once per month?}'').
Previous studies have performed separate analyses on the two questions \citep{koster_family_2018, simpson2020structural}. We examine both questions in a single model, examining the potential biases that shape the reports of social support. 
In this dataset the reports vary significantly across reporters, with some reporters nominating many ties while others much fewer.
It is, therefore, reasonable to have the priors on $\theta_m$ reflecting this. 
We can incorporate this insight by running the inference in two steps, where we first run \vimure \text{} with a weak prior that is the same for all reporters, while in a  second step we run \vimure \text{} with a prior proportional to the posterior mean of   $\theta_m$  inferred in the first step.
This allows us to obtain a wider range of reliabilities.
This is particularly important as we can distinguish possible exaggerators more distinctively than when using the same prior for all reporters.

\begin{figure}[!htb]
\centering
\includegraphics[width=0.95\linewidth]{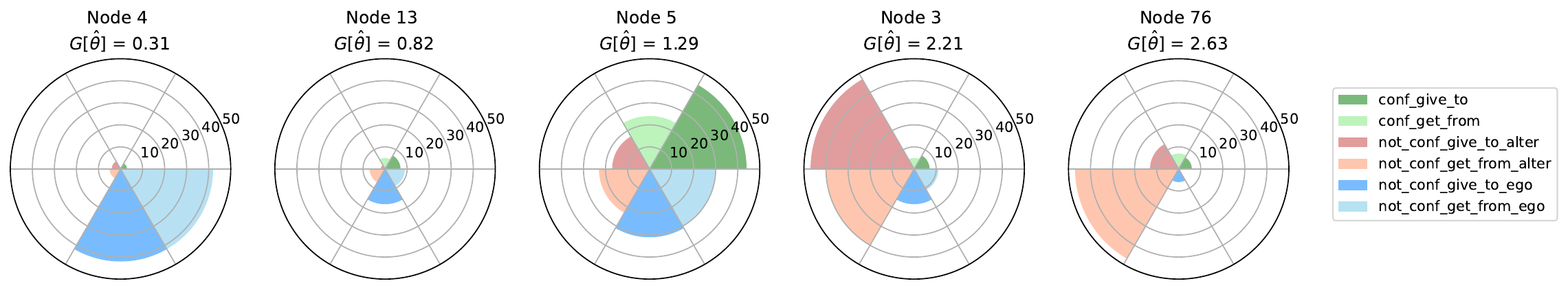}
\caption{Example of individual reliabilities. Pie plots show six different configurations for the reported ties (two per each direction of a tie): ties confirmed by both reporters (conf ``give to'', conf ``get from''); ties reported by $m$ but not confirmed by others (not conf ``give to'' (alter),  not conf ``get from'' (alter)); ties reported by others but not by $m$ (not conf ``give to'' (ego),  not conf ``get from'' (ego)).
Each plot is a different reporter; their estimated reliability $\hat{\theta}$ is printed on top. Each slice of the pie is one tie reported in one the six possible ways, represented by the colors. In this example we consider reporters from the Nicaragua dataset.}
\label{fig:pie}
\end{figure}  

As with our previous examples, applying \vimure\text{} produces estimates of a network that has properties (e.g., mean degree, reciprocity) that fall somewhere between the results of taking the union (which returns an incredibly dense network) and taking the intersection of the double-sampled ties (which returns an extremely sparse network).
See Table \Cref{tab:netsummary}, for a summary. 
Overall, mutuality was estimated to be $\eta_{est} = 0.610$ and reciprocity was $0.278$. 
These results align closely with the values found in the Karnataka data. 
Indeed, it is particularly close to the values for the ``Money'' tie type in the Karnataka data. 
The ``tangible support'' considered here would potentially be closest to that, or the ``Household Items'' tie type, which had higher values for both reciprocity and mutuality. 

In contrast to the Karnataka dataset, where survey design may contribute to under-reporting, the roster-based design may make it much easier for respondents to make many nominations. 
In the roster design in Nicaragua, informants reported approximately 25 alters for each prompt, substantially more than the average of 4 from the constrained name generators used in the Karnataka study (see \Cref{tab:netsummary} for full network statistics).
As can be seen in \Cref{fig:pie}, we observe reporters with low $\theta_m$ who were nominated by several others but nominated relatively few themselves (e.g., Nodes 4 and 13). On the opposite extreme, we see reporters with high $\theta_m$ who nominated many but these ties are not confirmed by the alters (e.g., Nodes 3 and 76). In between, we show an example of a reporter (Node 5) with intermediate value of $\theta_m$ who nominates several others and these ties are confirmed, but this person does not confirm several ties that others report. 
The distribution of reliability for reporters in this dataset can be seen in \Cref{fig:nicaraguaReliability}.

\section{Discussion and conclusions}

Self-report network data are an important resource for social scientists, but they are also susceptible to several types of reporter bias. 
Identifying the possible biases that structure self reports of social relationships remains an essential and open area of research for social network analysis. 
Failure to identify and account for such reporting biases may lead researchers to draw incorrect inferences.  
But how should social scientists go about investigating and treating reporting bias in measurements of social networks?
We provide a novel statistical solution to---and theoretical and empirical evidence of---this problem, with particular focus on two forms of bias.
First, we investigate and adjust for the general propensity of reporting balanced, reciprocal relationships (i.e., ``mutuality'').
Second, we advance statistical tools for more nuanced understanding of individuals' unique potential to misrepresent or misreport their relationships. 
Both of these forms of bias have the potential to add substantial ``noise'' to empirical representations of social networks, but the extent to which this is present and problematic has not yet been well established. 
While previous work has explored individual propensities to misreport their ties \citep[e.g.,][]{butts_network_2003,newman2018estimating, young_bayesian_2021}, there has been limited formal analysis of the impact that ``mutuality'' has on network inference \citep[but see,][]{redhead2021reliable}.

We have focused our attention on cases where multiple reporters are able to provide information on any given relationship, providing crucial empirical data that could be used to assess the biases under investigation here. 
In particular, we have considered ``double sampled'' relationships, where respondents are asked about their role both as giver and as receiver---a common technique that is used in social support network surveys. 
We have introduced a probabilistic modelling framework, \vimure, that provides a principled solution to these issues and aims to more appropriately capture the data generating process associated with name generator designs.
Within our framework, \vimure\text{} takes as input the potentially biased, imperfect survey responses and uses these to estimate a ``true'' latent network, as well as any individual biases and relationship-specific tendencies towards mutuality. 
The model estimates both a ground-truth $Y$ and $\theta$ which, in certain cases, can be interpreted as a reporter's reliability (while accounting for mutuality).

The model that we have introduced strongly departs from standard approaches for dealing with double-sampled network data in the social sciences, which involve researchers simply taking the union or the intersection of nominations.
Our approach also departs from existing network reconstruction methods, and advances a framework that is maximally flexible. 
To our knowledge, existing network reconstruction methods \citep[e.g.,][]{butts_network_2003, newman2018estimating, young_bayesian_2021} that are applicable to social networks focus on the single-sampled case--with the exception of \citet{redhead2021reliable}, which is applicable only to double-sampled networks.
While we have highlighted double-sampled network data, our framework can be readily used for many reporting sampling schemes.
A tie within a network could be reported on by any number of reporters, up to and including a full ``cognitive social structure'' design \citep{krackhardt_cognitive_1987}, where each respondent reports on all other ties in the network. 
Alongside this, the model is computationally efficient given the use of variational inference, as opposed to a Monte Carlo approach. 
Given this, our model can flexibly handle social network datasets of any realistic size, and can scale to large systems of tens of thousands of nodes by exploiting the sparsity of the network datasets. 

Results from our simulation experiments highlight that mutuality dramatically impacts inferred levels of reciprocity. 
Our results complement previous theoretical and empirical studies \citep{readymeasuring, redhead2021reliable}, and show that the simple deterministic rules of the union and intersection of nominations lead to over-estimates of reciprocity. 
Given this, we propose a simple heuristic that is based solely on the mutuality value inferred by the model, that can be used to select the most appropriate point-estimates from an estimated posterior distribution of $Y$.
Findings from our simulation experiments suggest that our approach results in networks that are somewhere between those produced by the union and the intersection.
Generally, our approach results in much lower levels of reciprocity than such deterministic aggregation, because we are appropriately accounting for mutuality. 

The importance of considering the core questions of (bias in) network representation---and the utility of \vimure---are most clearly demonstrated with our analyses of the empirical data from Karnataka \citep{banerjee_diffusion_2013} and Nicaragua \citep{koster_family_2018}.
These datasets result from two very different elicitation approaches, which carry with them different potential risks for bias. 
The data from Karnataka provide a case where a standard name generator approach was used on a partial sample of the network, and where an upper bound of four was placed on the number of ties that could be reported.
This design may likely increase the chances of under-reporting of ties.
In contrast, the data from Nicaragua were collected using a full roster-based design on the entire sample.
This approach may inflate the chances of over-reporting of ties.
In both empirical examples, the prospect of mutuality is salient, as reporters were asked about their roles as givers and receivers in direct succession, and there was no randomization of question order. 
The results of our empirical applications indicate the importance of mutuality in patterning reports within double-sampled designs.  
In Karnataka, mutuality values range from $\sim0.4$ to $\sim0.7$, and in Nicaragua they are $\sim 0.6$. 
This complements the findings from our simulation experiments, 
which show that when mutuality is high, taking either the union or the intersection will result in inflated reciprocity values (despite treating discordant responses in very different ways). 

Our findings highlight that mutuality is indeed high across a range of different relationship types, and thus the consequences of using these standard deterministic aggregation methods are obvious: a clear disparity between the resulting aggregated network and the ``true'' underlying network. 
How acute this issue is depends on the particular tie type, as we can see in the varying levels in the Karnataka data (where mutuality is lowest for relationships that may be seen as less balanced). 
\vimure\text{} provides a promising way forward here as it is able to measure and account for mutuality across a broad range of different sampling regimes. 

The empirical examples that we present further elucidate the varying ``reliabilities'' of reporters---over and above the general propensity to report mutually supportive, reciprocal ties. 
Importantly, the contrasting results found between the two sets of empirical data reveal general issues about sampling and elicitation that practitioners need to be cognisant of. 
Given the roster-based design used to collect the network data from Nicaragua, the average number of nominations that were given for each prompt was roughly 25. 
In contrast, given the upper limit of four ties that could be reported in the Karnataka design, the average number of nominations was much lower (two to three nominations for each prompt) for the various relationship types (see \Cref{tab:netsummary}).
Compounding this issue is the partial sampling procedure implemented in Karnataka.
The partial sample included $\sim46\%$ of the households and $\sim25\%$ of residents (including children).
Our findings suggest that when many nominations are of people who were not themselves reporters, there are considerable constraints on the ability to assess reliability. 
Moreover, our findings suggest that individuals who were named by many others are likely to be seen as ``unreliable'' (see \Cref{fig:thetaKarnataka}), in part because these individuals were constrained in their ability to name more than four individuals. 
Generally speaking, greater coverage of the network and prompts that result in more nominations will facilitate more precise estimation of individual ``reliabilities'' and, thus, more accurate network reconstruction. 

Several directions are possible for future improvements to \vimure. 
Our model specifies conditional probabilities, and thus relies on pseudo-likelihood estimation for inferring the parameters. 
A fruitful avenue for future research is to improve this approximation by characterizing a full joint distribution of a pair of ties \citep{contisciani2021JointCrep}. 
Doing this may potentially solve the problem of identifying a $\theta_m$ for samples with high mutuality and, most importantly, increase the accuracy of estimating posterior distributions for $Y$.
However, any improvement may come at the price of losing analytical tractability, or requiring less flexible approaches. 
We have focused here on capturing reciprocity, but this does not provide any guarantees of recovering automatically other network properties involving higher-order motifs, such as transitivity or triadic closure (see \Cref{tab:netsummary}). 
How to adapt our model to include them is open for future research.

Alongside this, there are several possibilities for future extensions of the \vimure \text{} framework that we have introduced here. 
First, \vimure \text{} takes as input a set of reported ties generated flexibly from a variety of data collection routines and we assumed that this is the main information known. 
However, if practitioners have access to additional information---such as covariates on nodes---this could be incorporated into the model. 
For instance, one can consider a suitable prior for the reliabilities $\theta_m$ that is based upon a given covariate. 
It would also be straightforward to extend our model by incorporating more informative priors about the ground-truth network, $Y$ (e.g., if the network had a known block structure). 
Second, many social networks are fundamentally multi-level, with nodes being nested within higher order units \citep[e.g., households, businesses, or schools;][]{lazega_multilevel_2015}.
Formulating an approach to flexibly incorporate multi-level networks further remains an open and important area for extending our \vimure \ framework. 
Finally, our focus has been on cases where social networks are static. 
Investigating how to effectively adapt our model for networks evolving over time is an open avenue for future work.\\

In sum, multiply-reported social network data clearly highlight the potential for biases in self-report data, but also provide that material which, if properly mobilized, can help to account for those biases. \vimure \text{} attempts to do so by explicitly modelling mutuality, the tendency of reporters of nominating the same individuals for both directions of a tie, and estimating a parameter $\theta_m$ for each reporter, which can be related to their tendency to report various amounts of ties. Inference is based on variational inference, leading to a fast algorithmic implementation that is scalable to large system sizes. 
Our study of the datasets from Karnataka and Nicaragua establishes that there is indeed important variation in reporters' ``reliability,'' and that people's reports seem to be driven in part by their normative expectation of relationships as balanced and reciprocal. We observe this high ``mutuality'' despite very different data elicitation approaches, and see that it varies based on the type of relationship being elicited. These findings demonstrate the value of employing a tool such as \vimure, as it can not only give crucial insights into how social relationships are understood by individuals, but can also provide a way to account for these individual and collective biases and arrive at a more appropriate representation of the network of interest. 
To facilitate its usage by practitioners, we provide an open source implementation of the code online. 
   
\section*{Acknowledgements}
\vspace{-0.1in}
This paper comes out of a collaboration funded by a UKRI Economic and Social Research Council Research Methods Development Grant (ES/V006495/1).
The authors thank the International Max Planck Research School for Intelligent Systems (IMPRS-IS) for supporting Martina Contisciani and Diego Baptista. 
Caterina De Bacco, Martina Contisciani and Hadiseh Safdari were supported by the Cyber Valley Research Fund. 
Daniel Redhead, Cody T. Ross, and Richard McElreath were supported by the Department of Human Behaviour, Ecology and Culture at the Max Planck Institute for Evolutionary Anthropology.

\section*{Data and Code}
All data and code used in this paper are available in the following public repository: \url{	
https://github.com/latentnetworks/vimure}.
The data that support the findings of this study are also openly available at the following locations: The Abdul Latif Jameel Poverty Action Lab Dataverse at \url{https://doi.org/10.7910/DVN/U3BIHX} and The Royal Society Open Science's Electronic Supplementary Material at \url{https://doi.org/10.1098/rsos.172159}. 

\bibliographystyle{rss}
\bibliography{bibliography}

\newcommand{\beginsupplement}{%
        \setcounter{table}{1}
        \renewcommand{\thetable}{S\arabic{table}}%
        \setcounter{figure}{1}
        \renewcommand{\thefigure}{S\arabic{figure}}%
        \setcounter{equation}{1}
        \renewcommand{\theequation}{S\arabic{equation}}
         \setcounter{section}{1}
        \renewcommand{\thesection}{S\arabic{section}}
 }

\clearpage
\beginsupplement

\section*{{Supporting Information (SI)}}
\subsection{Variational Inference}\label{sec:vi}
Variational Inference assumes approximate posterior distributions for the parameters \citep{blei2017variational}. We assume the mean-field variational family:

\be
{\tiny
q(\eta,\lambda,\theta,Y) =q(\eta;\nu^{shape},\nu^{rate}) \prod_{k,}q(\lambda_{k}^{};\phi_{k}^{shape},\phi_{k}^{rate})\, \prod_{m,}q(\theta_{m}^{};\gamma_{m}^{shape},\gamma_{m}^{rate})\prod_{i,j,} q(Y_{ij}^{};\rho_{ij}^{})\ , \\}
\label{eq:q}
\ee

\noindent where $q(\eta;\nu^{shape},\nu^{rate}),\,q(\lambda_{k}^{};\phi_{k}^{shape},\phi_{k}^{rate})$ and $q(\theta_{m}^{};\gamma_{m}^{shape},\gamma_{m}^{rate})$ are Gamma distributions, and $q(Y_{ij}^{};\rho_{ij}^{})$ is a Categorical distribution. The idea is to set the parameters of these distribution such that the KL divergence between this approximate posterior and the true posterior is minimized. This is equivalent to maximize the Evidence Lower Bound (ELBO), defined as:
\be
\text{ELBO}(q) = \Exp_{q}\rup{\log\mathcal{L}(\lambda,\theta, \eta, Y)} -\Exp_{q}\rup{\log q(\eta,\lambda,\theta,Y)} \ .
\ee
This could be done by taking the derivative w.r.t. each parameter of the variational distributions individually and setting them to zero. However, there is an alternative more direct way to extract the optimal parameters, which involves finding the complete conditionals of each variable \citep{blei2017variational}. This is defined as the conditional distribution of a given parameter, given all the remaining parameters and data.

\subsubsection{Auxiliary variables}
To facilitate inference we augment the model by introducing auxiliary variables that allow to make the model conditionally conjugate, i.e. all the complete conditionals are in the exponential family, which will eventually allow close-form and faster parameters' updates.
We define discrete latent auxiliary variables:
\be
z^{1}_{mk} \sim \pois(\theta_{m}^{}\lambda_{k}^{}), \quad z^{2}_{ijm} \sim \pois(\eta X_{jim}^{})  \quad  s.t. \quad  z^{1}_{mk}+z^{2}_{ijm}=X_{ijm}^{} \quad \vert \quad  Y_{ij}^{}=k\ .
\ee
With these new variables one can work with an augmented conditional distribution $P(X_{ijm}^{},Z|X_{jim}^{},Y_{ij}^{}=k, \lambda_k^{}, \theta_m^{}, \eta) $ which is factorized and thus easier to perform inference with. Notably, this joint conditional distribution is built such that its marginal recovers the original conditional distribution:
\be
P(X_{ijm}^{}|X_{jim}^{},Y_{ij}^{}=k, \lambda_k^{}, \theta_m^{}, \eta) =\sum_{z^{1}_{mk},z^{2}_{ijm}} \, P(X_{ijm}^{},Z|X_{jim}^{},Y_{ij}^{}=k, \lambda_k^{}, \theta_m^{}, \eta) \ .
\ee
This comes from the additive property of Poisson distributions: the sum of i.i.d. Poisson distributed variables is also Poisson distributed, with parameter the sum of the individual variables' parameters.
In fact:

{\small
\bea
P(X_{ijm}^{},Z|X_{jim}^{},Y_{ij}^{}=k, \lambda_k^{}, \theta_m^{}, \eta) &=& P(X_{ijm}^{}|Z,Y_{ij}^{}=k)\, P(Z |X_{jim}^{},\lambda_k^{}, \theta_m^{},\eta) \\
P(X_{ijm}^{}|Z,Y_{ij}^{}=k)&=&\delta\bup{z^{1}_{mk}+z^{2}_{ijm}-X_{ijm}^{}}^{Y_{ij,k}^{}} \\
P(Z |X_{jim}^{},\lambda_k^{}, \theta_m^{},\eta) &=& \pois(z^{1}_{mk};\theta_{m}^{}\lambda_{k}^{})\, \pois(z^{2}_{ijm};\eta X_{jim}^{})   \ ,
\eea}
where $\delta(\cdot)$ is the Dirac delta distribution. The variational distribution for $Z$ is a Multinomial distribution 
\be
q\bup{(z_{mkl}^1,z_{ijml}^2); X_{ijm}^{},\rup{\hat{z}_{mkl}^1, \hat{z}_{ijml}^2}} \ .
\ee
Assuming $z_{mkl}^1,z_{ijml}^2, \eta, \lambda_k^{}, \theta_m^{}$ and $Y_{ij}^{}$ are all independent, our goal is to infer the parameters of their variational distributions. When all the complete conditionals are in the exponential family, we can use the result of \citet{blei2017variational} that the natural parameters $\omega_{i}$ of the variational family satisfy:
\be\label{eqn:bestq}
\omega_{i}=\Exp_{q(y)} \kappa_{i}(y)\ ,
\ee
where $y$ is the parameter from the original posterior and $\kappa_{i}(y)$ is the natural parameter of the complete conditional. The expectation is with respect to the variational distribution $q(y)$.
 The natural parameters 	for a Gamma distribution $\Gam(\alpha, \beta)$ are $(\alpha-1, -\beta)$; for a Multinomial $\multi(n, [\log p_1, \dots, \log p_K])$ and a Categorical distribution $\cat([p_1, \dots, p_K])$ are $(\log p_1, \dots, \log p_K)$.
The full posterior distribution of the augmented system is:

\bea
P(Y, \lambda, \theta, \eta|X,Z) \nonumber
\eea
\vspace*{-2\baselineskip}
{\small
\bea
    &\text{    } & \nonumber\\
    &=&\prod_{i,j,,m,k} \rup{\bup{ \pois(z^{1}_{mk};\theta_{m}^{}\lambda_{k}^{})\, \pois(z^{2}_{ijm};\eta X_{jim}^{})}^{Y_{ij,k}^{}}   \delta\bup{z^{1}_{mk}+z^{2}_{ijm}-X_{ijm}^{}}^{Y_{ij,k}^{}}}\nonumber \\
    &&\times \; \underbrace{\Gam(\eta; c,d) \, \prod_{k,}  \Gam(\lambda_{k}^{};a_{k}, b_{k}) \, \prod_{m,} \; \Gam(\theta_{m}^{};\alpha_{m}, \beta_{m}) \, \prod_{i,j,}\cat(Y_{ij}^{};p_{ij}^{}) }_{\mbox{priors}} \ .
\eea}

\subsubsection{Update of $Z$} 
The complete conditional for  $Z$ is:

\bea
P(z^{1}_{mk},z^{2}_{ijm}|Z_{\setminus z^{1}_{mk},z^{2}_{ijm}}, X, Y, \lambda, \theta, \eta)  \nonumber
\eea
\vspace*{-3\baselineskip}
\bea
        & \text{      } & \nonumber \\
        & \propto & \delta\bup{z^{1}_{mk}+z^{2}_{ijm}-X_{ijm}^{}}^{Y_{ij,k}^{}}  \rup{ \f{X_{ijm}^{}!}{z^{1}_{mk}!\,z^{2}_{ijm}!}(\theta_{m}^{}\lambda_k^{})^{z^{1}_{mk}}\, (\eta X_{jim}^{})^{z^{2}_{ijm}}}^{Y_{ij,k}^{}}\\
        & \sim & \rup{\multi(X_{ijm}^{},\bar{z}_{ijm})}^{Y_{ij,k}^{}}\ ,
\eea

where $\bar{z}_{ijm}=(\theta_{m}^{}\lambda_{k}^{},\eta X_{jim}^{})$ are the parameters of the complete conditional.

Using \Cref{eqn:bestq} we get:
\bea
\log \hat{z}^{1}_{mk} &=& \Exp_{q(\theta_{m}^{})}\rup{\log \theta_{m}^{}} + \Exp_{q(\lambda_{k}^{})}\rup{\log \lambda_{k}^{}} \nonumber \\
&=& \Psi(\gamma_{m}^{shape})-\log \gamma_{m}^{rate}+\Psi(\phi_{k}^{shape})-\log \phi_{k}^{rate}  \ ,
\label{eq:z1}
\eea
where $\Psi(x)$ is the di-gamma function. Here we used $\Exp [\log x]= \Psi(a) -\log(b)$, valid for gamma-distributed variables.\\
Similarly:
\bea
\log \hat{z}^{2}_{ijm}&=& \Exp_{q(\eta)}\rup{\log \eta X_{jim}^{}}=\Psi(\nu^{shape}) -\log\nu^{rate} + \log X_{jim}^{}\ .
\label{eq:z2}
\eea

\subsubsection{Update of mutuality parameter $\eta$}
 The complete conditional for  $\eta$ is:
 \bea
P(\eta|X, Z, Y, \lambda, \theta) &\propto & \Gam(\eta;c,d)\prod_{i,j,,m,k}
 \pois(z^{2}_{ijm};\eta X_{jim}^{})^{Y_{ij,k}^{}} \nonumber \\
 &\propto & \eta^{c-1}e^{-d\eta}\prod_{i,j,,m,k}\rup{(\eta\, X_{jim}^{})^{z^{2}_{ijm}}e^{-\eta X_{jim}^{}}}^{Y_{ij,k}^{}}\\
 &\sim & \Gam\bup{c+ \sum_{i,j,,m,k}Y_{ij,k}^{} \,z^{2}_{ijm}, d+  \sum_{i,j,,m,k}Y_{ij,k}^{} X_{jim}^{}} \ .
\eea
Using \Cref{eqn:bestq} we get:
\bea
\nu^{shape} &=&c + \sum_{i,j,,m,k} \rho_{ij,k}^{} \, X_{ijm}^{} \,\hat{z}^{2}_{ijm}\\
\nu^{rate}&=&d+  \sum_{i,j,,m,k}\rho_{ij,k}^{}\, X_{jim}^{}\ .
\eea

\subsubsection{Update of $\lambda_{k}^{}$} 
The complete conditional for  $\lambda_{k}^{}$ is:
\bea
P(\lambda_{k}^{}|\lambda_{\setminus \lambda_k^{}}, X, Z, Y, \theta, \eta) &\propto & \Gam(\lambda_{k}^{};a_{k}, b_{k})\prod_{i,j,m}
 \pois(z^{1}_{mk};\theta_{m}^{}\lambda_{k}^{})^{Y_{ij,k}^{}} \nonumber \\
 &\propto & (\lambda_{k}^{})^{a_{k}-1}e^{-b_{k}\,\lambda_{k}^{}}\prod_{i,j,m}\rup{(\theta_{m}^{}\lambda_{k}^{})^{z^{1}_{mk}}e^{-\theta_{m}^{}\lambda_{k}^{}}}^{Y_{ij,k}^{}}\\
 &\sim & \Gam\bup{a_{k} + \sum_{i,j,m}Y_{ij,k}^{} \,z^{1}_{mk}, b_{k}+  \sum_{i,j,m}Y_{ij,k}^{} \,\theta_{m}^{}} \ .
\eea
Using \Cref{eqn:bestq} we get:
\bea
\phi_{k}^{shape} &=&a_{k} + \sum_{i,j,m} \rho_{ij,k}^{} \, X_{ijm}^{} \,\hat{z}^{1}_{mk}\\
\phi_{k}^{rate}&=&b_{k}+  \sum_{i,j,m}\rho_{ij,k}^{} \f{\gamma_{m}^{shape}}{\gamma_{m}^{rate}} \ .
\eea

\subsubsection{Update of reliabilities $\theta^{}_{m}$} 
The complete conditional for $\theta^{}_{m}$ is:

\bea
P(\theta^{}_{m}|\theta_{\setminus \theta_m^{}}, X, Z, Y, \lambda, \eta) &\propto & \Gam(\theta_{m}^{};\alpha_{m}, \beta_{m})\prod_{i,j,k}
 \pois(z^{1}_{mk};\theta_{m}^{}\lambda_{k}^{})^{Y_{ij,k}^{}} \nonumber \\
 &\propto & (\theta_{m}^{})^{\alpha_{m}-1}e^{-\beta_{m}\,\theta_{m}^{}}\prod_{i,j,k}\rup{(\theta_{m}^{}\lambda_{k}^{})^{z^{1}_{mk}}e^{-\theta_{m}^{}\lambda_{k}^{}}}^{Y_{ij,k}^{}}\\
 &\sim & \Gam\bup{\alpha_{m} + \sum_{i,j,k}Y_{ij,k}^{} \,z^{1}_{mk}, \beta_{m}+  \sum_{i,j,k}Y_{ij,k}^{} \,\lambda_{k}^{}} \ .
\eea
Using \Cref{eqn:bestq} we get:
\bea
\gamma_{m}^{shape} &=&\alpha_{m} + \sum_{i,j,k} \rho_{ij,k}^{} \, X_{ijm}^{} \,\hat{z}^{1}_{mk}\\
\gamma_{m}^{rate}&=&\beta_{m}+  \sum_{i,j,k} \rho_{ij,k}^{} \f{\phi_{k}^{shape}}{\phi_{k}^{rate}} \ .
\eea

\subsubsection{Update of true ties $Y_{ij}^{}$}

We have a complete conditional:

\bea
\hspace*{-1.2\baselineskip}
P(Y_{ij}^{}|Y_{\setminus Y_{ij}^{}},X,Z,\lambda,\theta,\eta) &\propto \nonumber
\eea
\vspace*{-3.7\baselineskip}

{\small
\bea
    &\text{   } & \nonumber\\
    &\propto& P(Y_{ij}^{})\,\prod_{k} \rup{\prod_{m} \bup{ \pois(z^{1}_{mk};\theta_{m}^{}\lambda_{k}^{})\, \pois(z^{2}_{ijm};\eta X_{jim}^{})}^{Y_{ij,k}^{}}  \delta\bup{z^{1}_{mk}+z^{2}_{ijm}-X_{ijm}^{}}^{Y_{ij,k}^{}}} \nonumber \\
    &\propto& \prod_{k} \rup{p_{ij,k}^{} \prod_{m} \bup{ \pois(z^{1}_{mk};\theta_{m}^{}\lambda_{k}^{})\, \pois(z^{2}_{ijm};\eta X_{jim}^{})}  \delta\bup{z^{1}_{mk}+z^{2}_{ijm}-X_{ijm}^{}}}^{Y_{ij,k}^{}} \\
    &\sim& \cat\bup{Y_{ij}^{};\rho_{ij}^{}}
\eea
}
\vspace*{-1.3\baselineskip}
\bea
    \rho_{ij,k}^{}&=&p_{ij,k}^{}\prod_{m}\pois(z^{1}_{mk};\theta_{m}^{}\lambda_{k}^{})\, \pois(z^{2}_{ijm};\eta X_{jim}^{})\delta\bup{z^{1}_{mk}+z^{2}_{ijm}-X_{ijm}^{}}\ .
\eea

Using \Cref{eqn:bestq} we get:

\bea
\hspace*{-1.2\baselineskip}
\rho_{ij,k}^{}  \nonumber
\eea
\vspace*{-2\baselineskip}
\bea
    &\propto \exp & \left\{\log p_{ij,k}^{}+\sum_{m} \bup{X_{ijm}^{} \hat{z}^{1}_{mk} \bup{\Exp_{q(\theta_m^{})}\rup{\log\theta_m^{}}+\Exp_{q(\lambda_k^{})} \rup{\log\lambda_k^{}}}} +\right. \nonumber \\
    && \quad \left.+ \sum_{m} X_{ijm}^{} \hat{z}^{2}_{ijm}\Exp_{q(\eta)} \rup{ \log \eta X_{jim}^{}} +\right. \nonumber \\
    &&\quad \left.-\f{\phi_{k}^{shape}}{\phi_{k}^{rate}}\sum_{m} \f{\gamma_{m}^{shape}}{\gamma_{m}^{rate}}   -\f{\nu^{shape}}{\nu^{rate}} \sum_{m}X_{jim}^{} +\right. \nonumber \\ 
    && \quad \left.-\sum_{m}\Exp_{q(z_{mk}^{1})}\rup{\log z^{1}_{mk}!}-\sum_{m}\Exp_{q(z_{ijm}^{2})}\rup{\log z^{2}_{ijm}!}\right\} \ .
\eea

Given that we have to normalize such that $\sum_{k}\rho_{ij,k}^{}=1$, we can shorten the equation by considering only terms that depend on $k$:
\bea
\rho_{ij,k}^{} &\propto& \exp \left\{\log  p_{ij,k}^{}+\sum_{m} \bup{X_{ijm}^{} \hat{z}^{1}_{mk} \Exp_{q(\lambda_k^{})} \rup{\log\lambda_k^{}}}  -\f{\phi_{k}^{shape}}{\phi_{k}^{rate}}\sum_{m} \f{\gamma_{m}^{shape}}{\gamma_{m}^{rate}}   
\right\} \ ,
\eea
where we used $z^{1}_{mk}=X_{ijm}^{}-z^{2}_{ijm}$ to remove terms that depend on $k$ only through $z^{1}_{mk}$.

\subsubsection{ELBO}
The Evidence Lower Bound (ELBO) is defined as:
\be
\text{ELBO}(q) = \Exp_{q}\rup{\log\mathcal{L}(\lambda,\theta, \eta, Y)} -\Exp_{q}\rup{\log q(\eta,\lambda,\theta,Y)} \ ,
\ee
where $\mathcal{L}(\lambda,\theta, \eta, Y)$ and $q(\eta,\lambda,\theta,Y)$ are given by Eq. (\ref{eqn:fullpost}) and \Cref{eq:q}. Thus,

\bea
\hspace*{-1.2\baselineskip}
\text{ELBO}(q) \quad \propto \nonumber
\eea
\vspace*{-2\baselineskip}

{\small
\bea
    &\text{ } & \nonumber\\
    &&  \Exp_{q} \rup{\sum_{i,j,l,m,k} Y_{ij,k}^{} \bup{X_{ijm}^{} \log(\theta_m^{} \lambda_k^{}+\eta X_{jim}^{}) - (\theta_m^{} \lambda_k^{}+\eta X_{jim}^{})}} + \nonumber \\ 
    &  + & \Exp_{q} \rup{\sum_{i,j,l,k} Y_{ij,k}^{}\log(p_{ij,k}^{})} \nonumber + \\
    &  + & \Exp_{q} \rup{c\log(\eta) -d\eta + \sum_{k,l} \bup{a_{kl}\log(\lambda_k^{}) - b_{kl}\lambda_k^{}} + \sum_{m,l} \bup{\alpha_{ml}\log(\theta_m^{}) - \beta_{ml}\theta_m^{}}} + \nonumber \\
    & - & \Exp_{q} \rup{\sum_{i,j,l,k} \bup{Y_{ij,k}^{}\log(\rho_{ij,k}^{})} + \nu^{shape}\log(\nu^{rate}) - \log(\Gamma(\nu^{shape})) + \nu^{shape}\log(\eta) -\nu^{rate}\eta} +\nonumber \\ 
    &  - & \Exp_{q} \rup{ \sum_{k,l} \bup{\phi_{k}^{shape}\log(\phi_{k}^{rate}) - \log(\Gamma(\phi_{k}^{shape})) + \phi_{k}^{shape}\log(\lambda_{k}^{}) -\phi_{k}^{rate}\lambda_{k}^{}}} +\nonumber\\
    &  - & \Exp_{q} \rup{ \sum_{m,l} \bup{\gamma_{m}^{shape}\log(\gamma_{m}^{rate}) - \log(\Gamma(\gamma_{m}^{shape})) + \gamma_{m}^{shape}\log(\theta_{m}^{}) -\gamma_{m}^{rate}\theta_{m}^{}}} \ ,
\eea}
where the proportionality is because we are omitting constant terms. Remember:
\bea
 \Exp_{q}  \rup{Y_{ij,k}^{}} &=& \rho_{ij,k}^{} \\
 \Exp_{q}  \rup{\theta_{m}^{}} &=& \f{\gamma_{m}^{shape}}{\gamma_{m}^{rate}} \\
 \Exp_{q}  \rup{\lambda_{k}^{}} &=& \f{\phi_{k}^{shape}}{\phi_{k}^{rate}} \\
 \Exp_{q}  \rup{\eta} &=&  \f{\nu^{shape}}{\nu^{rate}} \\
 \Exp_{q}  \rup{\log(\theta_{m}^{})} &=& \Psi(\gamma_{m}^{shape}) -\log(\gamma_{m}^{rate})\\
 \Exp_{q}  \rup{\log(\lambda_{k}^{})} &=& \Psi(\phi_{k}^{shape}) -\log(\phi_{k}^{rate})\\
 \Exp_{q}  \rup{\log(\eta)} &=& \Psi(\nu^{shape}) -\log(\nu^{rate}) \ .
 \eea
Moreover, we use the approximation:
\be
 \Exp_{q}  \rup{ \log(\theta_m^{} \lambda_k^{}+\eta X_{jim}^{})} =  \Exp_{q}  \rup{ \log(\theta_m^{} \lambda_k^{})} + \Exp_{q}  \rup{ \log(\eta X_{jim}^{})} \ ,
\ee
where $ \Exp_{q}  \rup{ \log(\theta_m^{} \lambda_k^{})}$ and $\Exp_{q}  \rup{ \log(\eta X_{jim}^{})}$ are defined in Eq. (\ref{eq:z1}) and  Eq. (\ref{eq:z2}). Then, the ELBO is given by:

\bea
\hspace*{-1.2\baselineskip}
\text{ELBO}(q) \quad \propto \nonumber
\eea
\vspace*{-7\baselineskip}

{\small
\bea
    &\text{ } & \nonumber\\
    && \sum_{i,j,l,m,k}  \rup{\rho_{ij,k}^{} \bup{X_{ijm}^{} \bup{\Exp_{q}  \rup{ \log(\theta_m^{} \lambda_k^{})} + \Exp_{q}  \rup{ \log(\eta X_{jim}^{})}} - \bup{ \f{\gamma_{m}^{shape}}{\gamma_{m}^{rate}}\f{\phi_{k}^{shape}}{\phi_{k}^{rate}}+ \f{\nu^{shape}}{\nu^{rate}} X_{jim}^{}}}} + \nonumber \\ 
    && + \Psi(\nu^{shape})(c - \nu^{shape}) + \log(\Gamma(\nu^{shape})) - c\log(\nu^{rate}) + \nu^{shape}\bup{1-\f{d}{\nu^{rate}}} + \nonumber  \\
    && + \sum_{k,l} \rup{ \Psi(\phi_{k}^{shape})(a_{kl} - \phi_{k}^{shape}) + \log(\Gamma(\phi_{k}^{shape})) - a_{kl}\log(\phi_{k}^{rate}) + \phi_{k}^{shape}\bup{1-\f{b_{kl}}{\phi_{k}^{rate}}}} +\nonumber  \\
    && + \sum_{m,l} \rup{ \Psi(\gamma_{m}^{shape})(\alpha_{ml} - \gamma_{m}^{shape}) + \log(\Gamma(\gamma_{m}^{shape})) - \alpha_{ml}\log(\gamma_{m}^{rate}) + \gamma_{m}^{shape}\bup{1-\f{\beta_{ml}}{\gamma_{m}^{rate}}}} + \nonumber \\
    && + \sum_{i,j,l,k} \rup{ \rho_{ij,k}^{} \bup{\log( p_{ij,k}^{}) - \log( \rho_{ij,k}^{})}} \ . 
\eea}

\subsection{Deriving the expected value of the marginal distribution}\label{sec:meanMarginal} 
\bea
\Exp \rup{X_{ijm}^{} \vert Y_{ij}^{}=k} &=& \mu_{ijm}^{} = \sum_{X_{ijm}^{},X_{jim}^{}} X_{ijm}^{} \, P(X_{ijm}^{},X_{jim}^{}|\Theta)  \nonumber \\
 &=& \sum_{X_{jim}^{}} P(X_{jim}^{}|\Theta)\,\sum_{X_{ijm}^{}} X_{ijm}^{}  \,P(X_{ijm}^{}|X_{jim}^{},\Theta) \nonumber \\
&=& \sum_{X_{jim}^{}} P(X_{jim}^{}|\Theta)\, \rup{\theta_m^{}\lambda_{k \vert Y_{ij}^{}=k}^{}+\eta \, X_{jim}^{}} \nonumber \\
&=&\theta_m^{}\lambda_{k \vert Y_{ij}^{}=k}^{}+\eta \, \sum_{X_{jim}^{}}  X_{jim}^{}\, P(X_{jim}^{}|\Theta) \nonumber \\
&=&\theta_m^{}\lambda_{k \vert Y_{ij}^{}=k}^{} +\eta \, \mu_{jim}^{}  \nonumber \\
&=&   \theta_m^{}\lambda_{k \vert Y_{ij}^{}=k}^{} +\eta \,\bup{\theta_m^{}\lambda_{k \vert Y_{ji}^{}=k}^{} +\eta \, \mu_{ijm}^{}} \quad.
\eea 
Solving for $\mu_{ijm}^{}$ yields:
\bea
\mu_{ijm}^{} \, \bup{1-\eta^{2}} &=&  \bup{ \theta_m^{}\lambda_{k \vert Y_{ij}^{}=k}^{}+ \eta \,  \theta_m^{}\lambda_{k \vert Y_{ji}^{}=k}^{}} \quad,
\eea
which implies:
\be \label{SIeqn:mij}
\mu_{ijm}^{} =\f{ \theta_m^{}\lambda_{k \vert Y_{ij}^{}=k}^{}+ \eta \,  \theta_m^{}\lambda_{k \vert Y_{ji}^{}=k}^{}}{ \bup{1- \eta^{2}} }\quad.
\ee

\subsection{Synthetic data generation} \label{sec:syntheticgeneration}
The synthetic networks used in the analysis are of three types and represent different scenarios: networks with fixed reliability and a fraction of over-reporters (a), networks with fixed reliability and a fraction of under-reporters (b), and networks with randomly generated reliabilities $\theta_m^{}$ (c). In these experiments, the true unobserved directed and binary networks $Y$ are generated by using either a stochastic block model (a), (b), or a degree-corrected stochastic block model (c). In all cases, the networks have $N = 100$ nodes, $C= 2$ non-overlapping communities of equal size, and average degree $\langle k \rangle=10$. In the scenario (c), the degree sequences are extracted from a power law distribution with exponents equal to $2$ and $2.5$ for the in-degree and the out-degree respectively. The networks follow an assortative structure and tie probabilities are $p_{in} = \langle k \rangle C / N$ and $p_{out} = 0.1\, p_{in}$, for within-group and between-group ties. To generate the observed directed and weighted double-sampled networks $X$, we need additional parameters to encode the individual reliability $\theta$, the mutuality effect $\eta$ and the contribution of the underlying data $Y$ in determining $X$, i.e. $\lambda$. Then, the networks $X$ are generated in a two-step routine:
\begin{enumerate}
\item Select with a coin-flip one direction, $(i,j)$ or $(j,i)$. Say we select $(i, j)$.         
\item Sample $X_{ijm}^{}$ from the marginal $P(X_{ijm}^{}|\Theta)=\pois(\mu_{ijm}^{})$, where $$\mu_{ijm}^{} = \f{\theta_m^{} \lambda_{k \vert Y_{ij}^{}=k}^{} + \eta \, \theta_m^{} \lambda_{k \vert Y_{ji}^{}=k}^{} }{1-\eta^2}$$ is the mean of the marginal such that it is consistent with the joint and the conditional distributions. For further details see \Cref{sec:meanMarginal}.
\item Sample $X_{jim}^{}$ from the conditional $P(X_{jim}^{}\vert X_{ijm}^{},Y_{ji}^{}=k, \lambda_k^{}, \theta_m^{}, \eta)=\pois(\theta_m^{} \lambda_k^{} + \eta X_{ijm}^{})$ using the previously extracted value of $X_{ijm}^{}$. 
\end{enumerate} 
We use $M=100$ reporters and a reporter $m$ only reports ties involving her, i.e. $R^{}_{ijm}=0$ when $m \not \in \ccup{i,j}$. In this equation, $X$ is weighted while $Y$ is binary so $\lambda^{} = [\lambda_0^{}, \lambda_1^{}]$. Furthermore, in these experiments we consider single-layer networks, i.e. $L=1$.

In scenarios (a) and (b) we vary the fraction of non-reliable individuals $\theta_{ratio} \in [0, 0.1, \dots, 0.5]$, where we avoid values bigger than 0.5 because they would result in unlikely and non-meaningful regimes. The reliability for this fraction of people is $\theta_m=50$ and $\theta_m=0.5$ for the over- and under-reporters respectively. All the other subjects have $\theta_m=1$ and entries of $X$ determined deterministically by using \Cref{eqn:meancond}. In these settings $\lambda = [0.01, 1.0]$, while $\eta \in \{0, 0.2\}$ and $\eta \in \{0, 0.6\}$ in scenarios (a) and (b) respectively. The different values of the mutuality parameter are used to provide comparable behaviours in the two regimes that are different because of the magnitudes of the mean of the Poisson. 

In the third scenario, $\theta_m$ is sampled from a Gamma distribution with $\alpha=\beta=2$, the mutuality effect is given and varies as $\eta  \in \{0, 0.2, 0.6\}$, while $\lambda_0 = 0.01$ and $\lambda_1 = \lambda_0^{} + \lambda_{diff}$ with $\lambda_{diff} \in [0.1, 0.2, \dots, 1]$.

\subsection{Recoverability of latent network properties in synthetic experiments}\label{sec:rho_threshold}

\subsubsection{Planted mutuality vs inferred mutuality}

\begin{figure}[!hbt]
	\centering
	\includegraphics[trim={0 0 0 0.9cm},clip,width=0.75\textwidth]{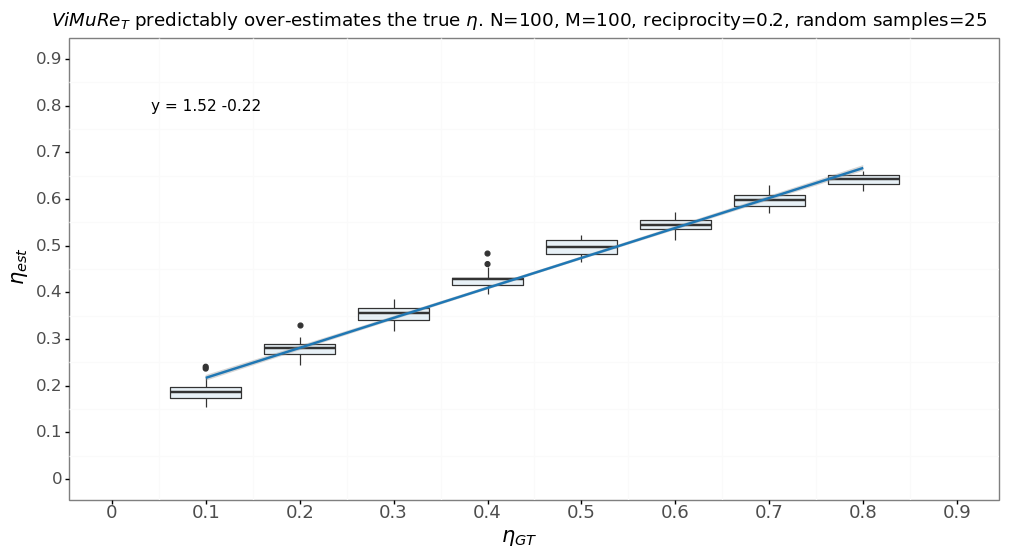}
	\caption{Synthetic networks with $N=100$ nodes and $M=100$ reporters generated with the benchmark generative model  with $\lambda_{1}- \lambda_{0} = 1.0$ and planted reciprocity values around $\approx 0.2$ on the ground truth networks. The plot compares the distributions of mutuality values as recovered by \vimure$_T$ ($\eta_{est}$) to the true planted mutuality on $X$, $\eta_{GT}$.}
	\label{fig:synTrueVsEstEta}
\end{figure}

From the experiments with synthetic networks, we observe that \vimure$_T$ is capable at distinguishing between the scenarios of low and high mutuality on respondents' data, $X$.
In fact, the mutuality inferred by our model ($\eta_{est}$) is highly correlated with the planted ground truth mutuality ($\eta_{GT}$), as it can be seen in Figure \ref{fig:synTrueVsEstEta}.
Mutuality inferred by our model is therefore reliable, requiring at most a downward adjustment when interpreting network characteristics.
\begin{figure}[!htbp]
\centering
\includegraphics[width=0.95\linewidth]{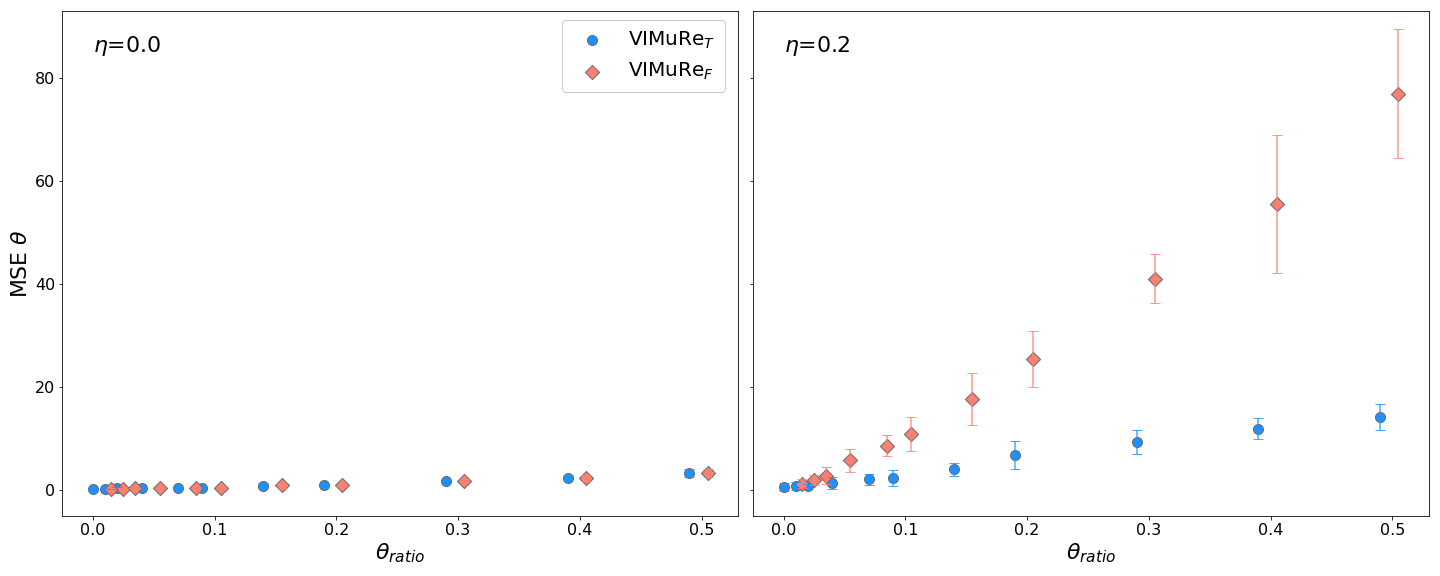}
\includegraphics[width=0.95\linewidth]{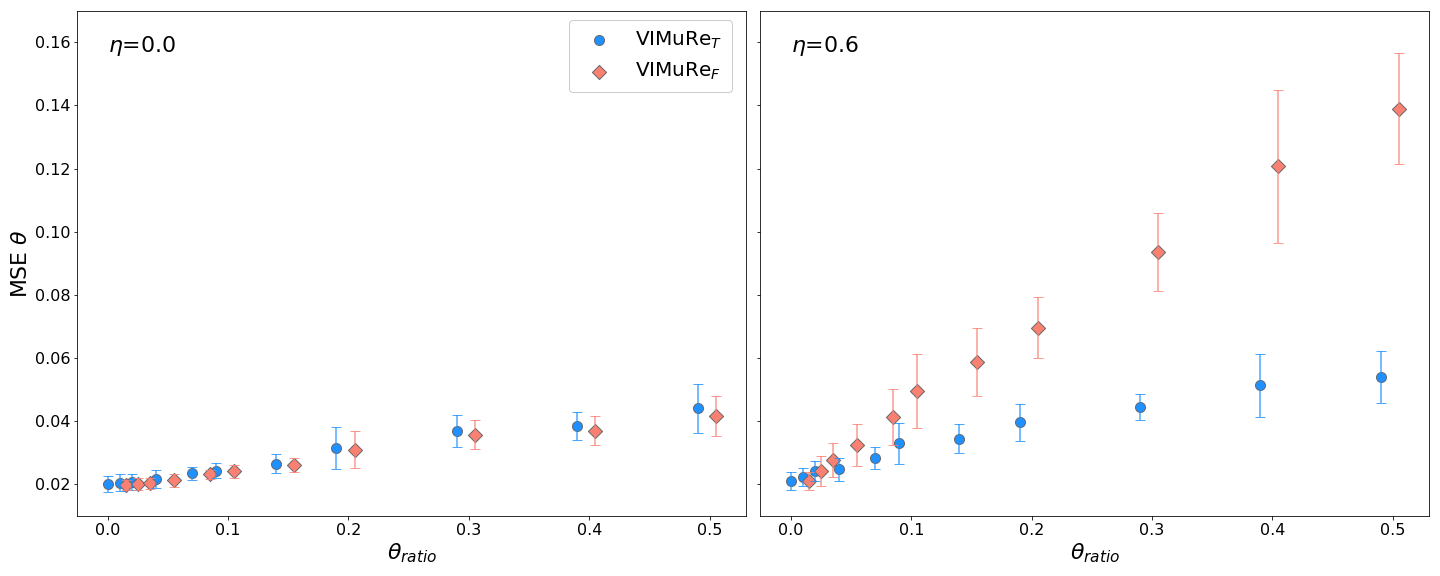}
\caption{Synthetic networks with over- or under-reporters: estimating reliabilities $\theta$. Synthetic networks with $N=100$ nodes and $M=100$ reporters generated with the benchmark generative model proposed above by varying the fraction $\theta_{ratio}$ of over-reporters (top) or under-reporters (bottom). The two columns represent networks generated  without (left) and with (right) the mutuality effect $\eta$. The results are averages and standard deviations over ten independent synthetic networks, and the accuracy over the reliabilities $\theta$ is measured with the Mean Squared Error. Values close to 0 indicate perfect recovery.}
\label{fig:synExaUnderRel}
\end{figure}  

\begin{figure}[!htb]
\centering
\includegraphics[width=0.75\textwidth]{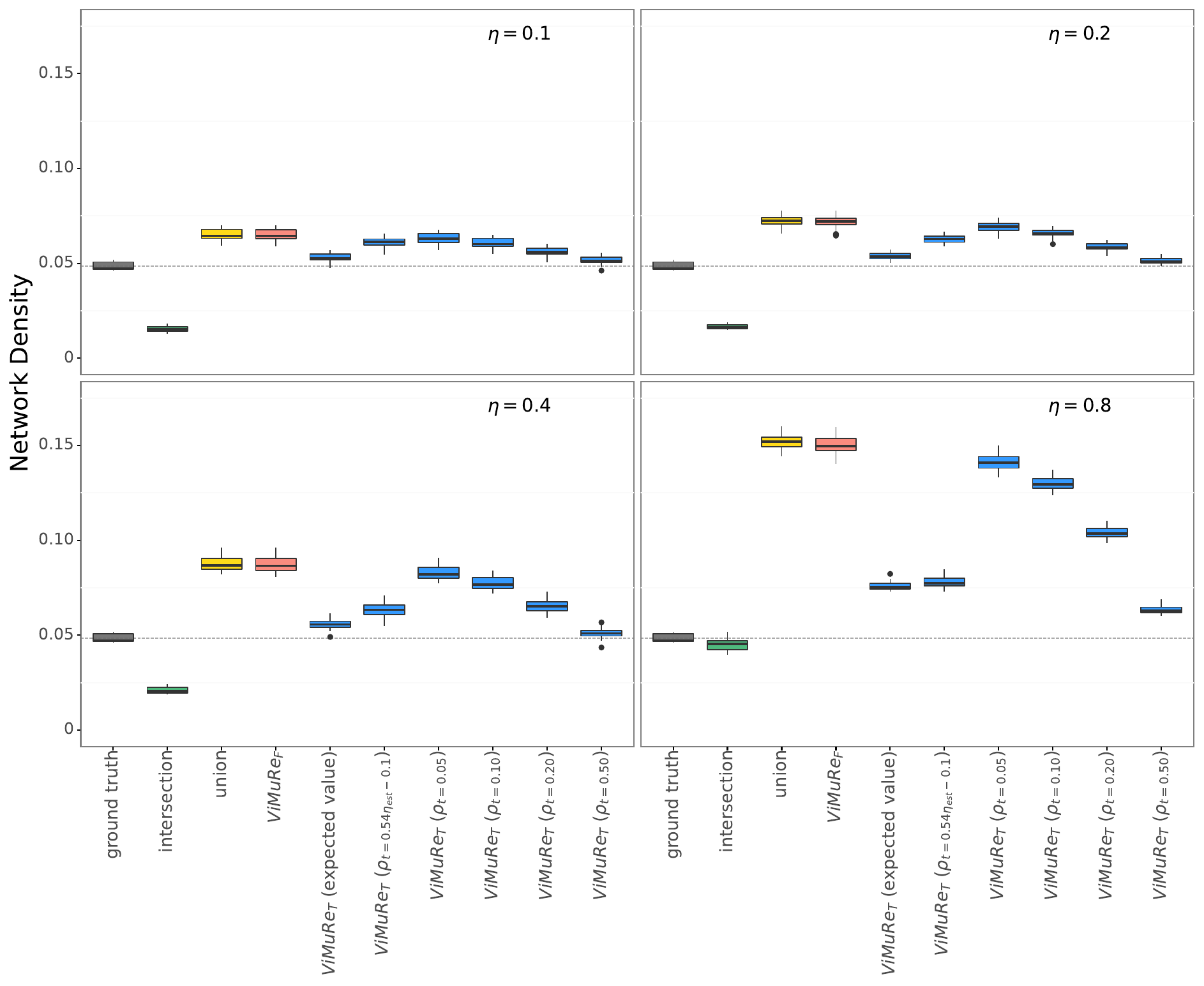}
\caption{Synthetic networks with $N=100$ nodes and $M=100$ reporters generated with the benchmark generative model with $\lambda_{1}- \lambda_{0} = 1.0$ and planted reciprocity values around $\approx 0.2$ on the ground truth network. The four sub-plots represent networks generated with low (top left), medium (top right) to increasingly high (bottom left and right) mutuality effect $\eta$. The box plots are distributions of the network density on the $Y$, over twenty samples of synthetic networks.}
\label{fig:synBothYDensity}
\end{figure}  

\begin{figure}[!htb]
\centering
\includegraphics[width=0.85\textwidth]{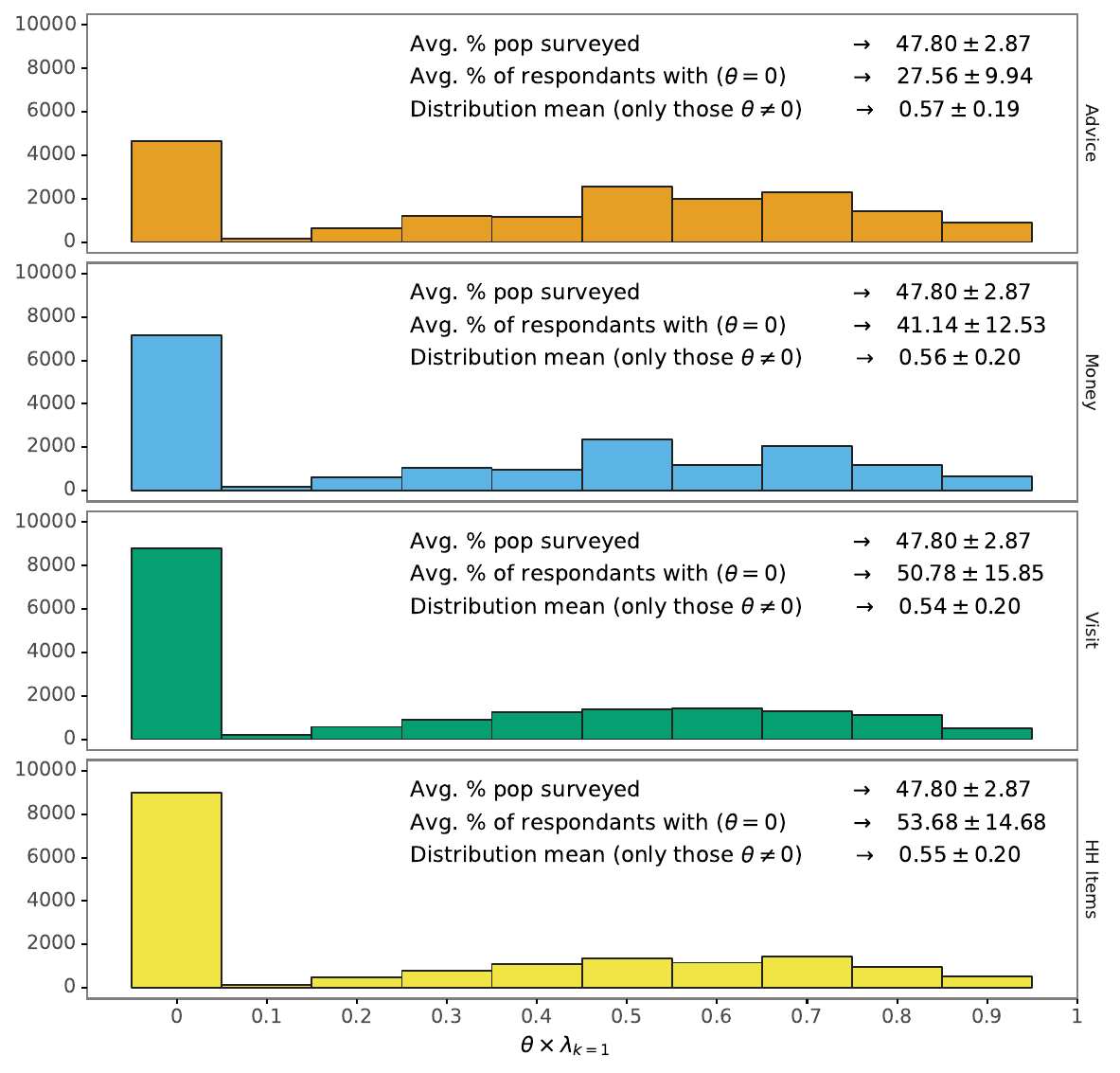}
\caption{Distribution of reliability estimates across different tie types in Karnataka networks.}
\label{fig:karnatakaReliability}
\end{figure}  

\begin{figure}[!htb]
\centering
\includegraphics[width=0.85\linewidth]{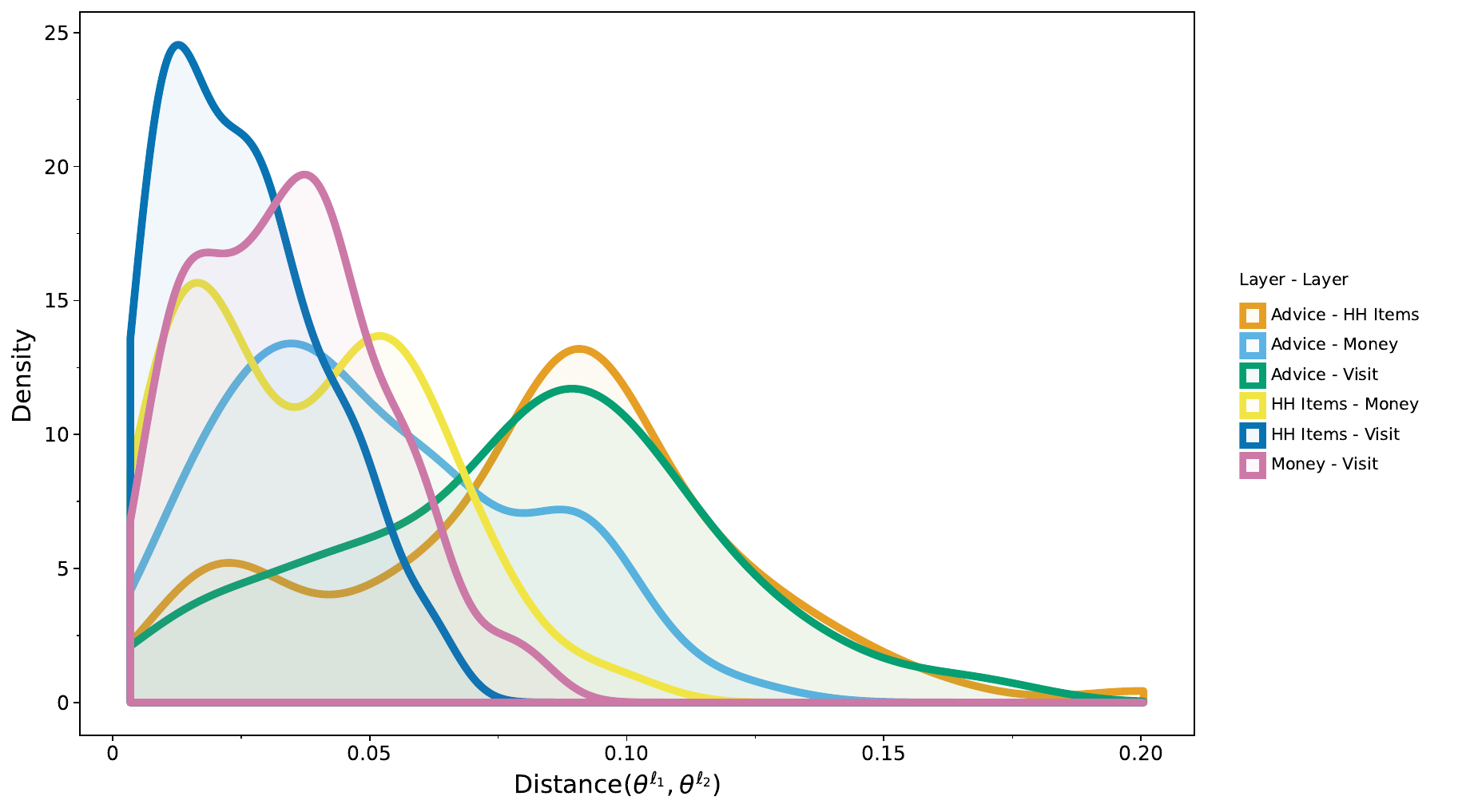}
\caption{
Wasserstein distance of reporters' reliabilities between different tie types in Karnataka networks, measured by the Wasserstein metric.}
\label{fig:thetaKarnataka}
\end{figure}  

\begin{figure}[!htb]
\centering
\includegraphics[width=0.8\linewidth]{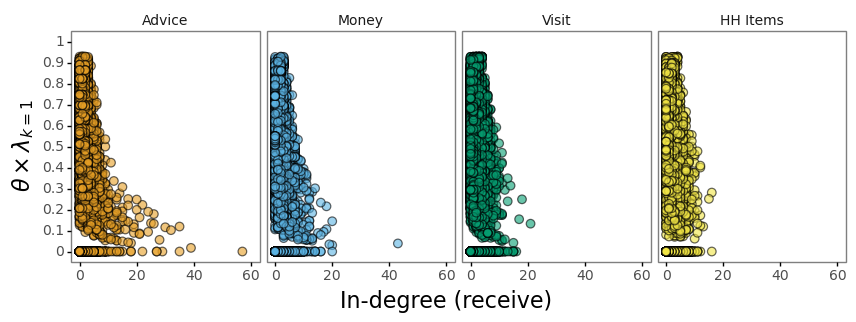}
\caption{In-degree (receive) vs individual reliability in Karnataka networks per tie type.}
\label{fig:indegthetaKarnataka}
\end{figure}  

\begin{figure}[!htb]
\centering
\includegraphics[width=0.65\textwidth]{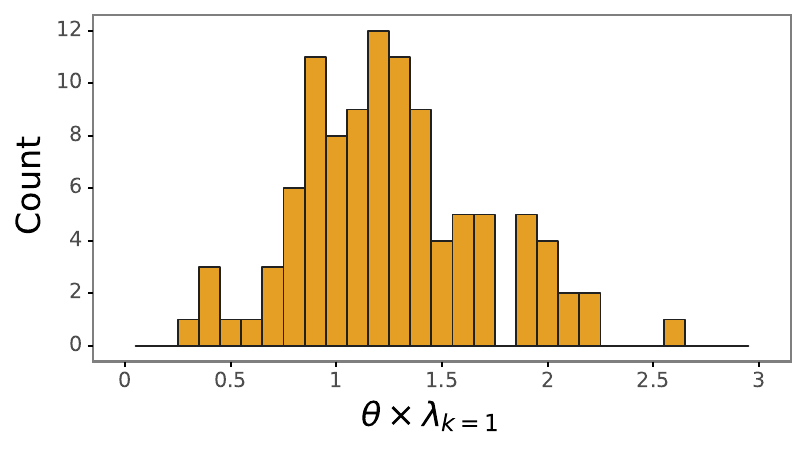}
\caption{Distribution of reliability estimates of respondent nodes in the Nicaragua dataset, obtained after the 2-step process.}
\label{fig:nicaraguaReliability}
\end{figure}  

\clearpage
\renewcommand{\arraystretch}{1.1}
\setlength{\tabcolsep}{2pt}


\begin{small}
\begin{longtable}{@{\extracolsep{\fill}}p{1mm}l@{\hspace{0.025\linewidth}}r>{$\pm}r<{$}r@{\hspace{0.025\linewidth}}r>{$\pm}r<{$}r@{\hspace{0.025\linewidth}}r>{$\pm}r<{$}r@{\hspace{0.025\linewidth}}r>{$\pm}r<{$}r@{\extracolsep{\fill}}}
     \multicolumn{2}{l}{\textbf{Networks}} & \multicolumn{3}{c}{\textbf{Edges \text{  }}} &    \multicolumn{3}{c}{\textbf{Mean Degree}}  &  \multicolumn{3}{c}{\textbf{Transitivity  }} &  \multicolumn{3}{c}{\textbf{Reciprocity  }} \\
    & & (mean & & std) & (mean & & std) & (mean & & std) & (mean & & std)\\
    \midrule
    
    \multicolumn{2}{l}{\textbf{Karnataka}} \\ 
    \multicolumn{2}{l}{(75 villages)} \\ \addlinespace[8pt]
    
    & Advice receive &   $ 243.04 $ & & $ 83.92$ &  $ 1.21$ && $0.16$ &  $ 0.14$ && $0.06$ &  $ 0.20$ && $0.08$ \\
    & Advice give &     $ 195.96 $ &&  $  71.39$ &  $ 1.05$ && $0.11$ &  $ 0.04$ && $0.02$ &  $ 0.22$ && $0.09$ \\
    & Advice (intersection) & $ 44.69 $ & & $  23.51$ &  $ 0.20$ && $0.09$ &  $ 0.06$ && $0.16$ &  $ 0.45$ && $0.18$ \\ 
    & Advice (union) &  $ 394.31 $ & & $ 137.06$ &  $ 1.74$ && $0.26$ &  $ 0.11$ && $0.04$ &  $ 0.50$ && $0.11$\\ 
    & \bf{\vimure} &  $\mathbf{343.57}$ & & $\mathbf{117.39}$ &  $\mathbf{1.52}$ && $\mathbf{0.21}$ &  $ \mathbf{0.11}$ && $\mathbf{0.03}$ &  $\mathbf{0.39}$ && $\mathbf{0.07}$ \\\addlinespace[8pt]
    
    & Money receive &  $ 249.03 $ & & $ 102.08$ &  $ 1.28$ && $0.20$ &  $ 0.13$ && $0.06$ &  $ 0.15$ && $0.05$ \\
    & Money give &   $ 217.53 $ &&  $ 88.44$ &  $ 1.18$ && $0.17$ &  $ 0.05$ && $0.03$ &  $ 0.16$ && $0.05$ \\
    & Money (intersection) &  $ 37.43 $ &&  $ 19.56$ &  $ 0.17$ && $0.07$ &  $ 0.06$ && $0.14$ &  $ 0.58$ && $0.17$ \\ 
    & Money (union) &  $ 429.13 $ &&  $ 174.32$ &  $ 1.88$ && $0.39$ &  $ 0.11$ && $0.05$ &  $ 0.64$ && $0.09$ \\ 
    & \bf{\vimure} &  $\mathbf{322.17} $ & & $ \mathbf{135.20}$ &  $\mathbf{1.40}$ && $\mathbf{0.29}$ &  $ \mathbf{0.09}$ && $\mathbf{0.04}$ &  $\mathbf{0.43}$ && $\mathbf{0.08}$ \\ \addlinespace[8pt]

    & Visit go &  $ 344.25 $ & & $ 142.18$ &  $ 1.64$ && $0.30$ &  $ 0.16$& & $0.06$ &  $ 0.19$ && $0.05$ \\
    & Visit come &  $ 328.99 $ &&  $ 133.76$ &  $ 1.58$ && $0.27$ &  $ 0.07$ && $0.03$ &  $ 0.19$ && $0.05$ \\
    & Visit (intersection) & $ 63.97 $ && $ 31.70$ &  $ 0.28$ && $0.11$ &  $ 0.14$ && $0.17$ &  $ 0.70$ && $0.16$ \\ 
    & Visit (union) &  $ 609.27 $ & & $ 249.48$ &  $ 2.66$ && $0.59$ &  $ 0.13$ && $0.05$ &  $ 0.76$ && $0.10$ \\
    & \bf{\vimure} &  $\mathbf{389.60}$ &&  $\mathbf{189.50}$ &  $\mathbf{1.69}$ && $\mathbf{0.49}$ &  $ \mathbf{0.11}$ && $\mathbf{0.04}$ & $\mathbf{0.47}$ && $\mathbf{0.07}$\\ \addlinespace[8pt]

    & Household items receive &  $ 289.08 $ &&  $ 118.49$ &  $ 1.45$ && $0.24$ &  $ 0.19$ && $0.07$ &  $ 0.20$ && $0.05$ \\
    & Household items give &  $ 281.43 $ &&  $ 111.24$ &  $ 1.41$ && $0.22$ &  $ 0.07$ && $0.03$ &  $ 0.18$ && $0.05$ \\
    & Household items (intersection) & $ 54.67$ &&  $29.97$ &  $ 0.24$ && $0.11$ &  $ 0.19$ && $0.22$ &  $ 0.72$ && $0.16$ \\
    & Household items (union) &  $ 515.84$ & & $204.45$ &  $ 2.26$ && $0.46$ &  $ 0.14$ && $0.05$ &  $ 0.74$ && $0.12$ \\
    & \bf{\vimure} &  $\mathbf{350.68}$ &&  $\mathbf{159.07}$ &  $\mathbf{1.53}$ && $\mathbf{0.38}$ &  $ \mathbf{0.11}$ && $\mathbf{0.05}$ &  $\mathbf{0.48}$ && $\mathbf{0.08}$ \\ \\
    
    \multicolumn{2}{l}{\textbf{Nicaragua}} \\ 
    \multicolumn{2}{l}{(108 nodes, 106 respondents)}\\ \addlinespace[8pt]
    
    & Incoming  &          \multicolumn{3}{c}{$2595$}  &  $24.03$ & & $12.12$ & \multicolumn{3}{c}{$0.40$}&        \multicolumn{3}{c}{$0.46$}\\
    & Outgoing &         \multicolumn{3}{c}{$2958$} & $ 27.39$ & & $16.34$ &\multicolumn{3}{c}{$0.40$} &       \multicolumn{3}{c}{$0.55$}\\
    & Intersection  &              \multicolumn{3}{c}{$1422$} &  $ 13.17$ && $7.69$ & \multicolumn{3}{c}{$0.38$} &        \multicolumn{3}{c}{$0.71$}\\
    & Union        &             \multicolumn{3}{c}{$4131$} & $ 38.25$ && $16.82$ & \multicolumn{3}{c}{$0.52$}     &   \multicolumn{3}{c}{$0.76$}\\
    &  \bf{\vimure} &     \multicolumn{3}{c}{$\mathbf{1517}$} &  $\mathbf{14.05}$ && $\mathbf{7.18}$ & \multicolumn{3}{c}{$\mathbf{0.16}$} &  \multicolumn{3}{c}{$\mathbf{0.11}$} \\

    \bottomrule
    \caption{Summary statistics of original network layers, baseline aggregations and networks inferred by \vimure.}
    \label{tab:netsummary}
\end{longtable}
\end{small}

\end{document}